\documentclass[preprint,prc,aps,showpacs,showkeys,groupedaddress,floatfix]{revtex4-1}
\usepackage{epsfig}
\usepackage{dcolumn}
\usepackage{bm}
\usepackage[latin5]{inputenc}
\usepackage{graphics}
\usepackage{graphicx}
\usepackage{epsfig}
\usepackage{amssymb}
\usepackage{amsmath}

\begin{document}
\title{Analytical solution of the Klein Gordon equation with a Multi-parameter q-Deformed Woods-Saxon Type Potential}
\author{B.C. L\"{u}tf\"{u}o\u{g}lu}
\affiliation{Department of Physics, Akdeniz University, 07058 Antalya, Turkey}

\affiliation{Department of Physics, Faculty of Science, University of Hradec Kr\'{a}lov\'{e}, Rokitansk\'{e}ho 62, 500\,03 Hradec Kr\'{a}lov\'{e}, Czechia}

\author{A.N. Ikot}
\affiliation{Theoretical Physics Group,Department of Physics, University of Port Harcourt, P M B 5323
Choba, Port Harcourt, Nigeria}
\author{E.O. Chukwocha}
\affiliation{Theoretical Physics Group,Department of Physics, University of Port Harcourt, P M B 5323
Choba, Port Harcourt, Nigeria}
\author{F.E. Bazuaye}
\affiliation{Department of Mathematics, University of Port Harcourt, Nigeria}
\date{\today}
\begin{abstract}
In this manuscript, we present analytical solution of the Klein-Gordon equation with the multi-parameter q-deformed Woods-Saxon type potential energy under the spin symmetric limit in $(1+1)$ dimension. In the scattering case, we obtain the reflection and transmission probabilities and prove the conservation of the total probability. Moreover, we analyze the correlation between the potential parameters with the reflection and transmission probabilities. In the bound state case, we use the continuity conditions and derive a quantization scheme.  To confirm our results numerically, in both cases we randomly assign values to the potential parameters and find numerical results by using the Newton Raphson method.
\end{abstract}

 \maketitle 

\section{Introduction}\label{sec1:intro}
Klein Gordon (KG) equation is a second order relativistic wave equation in space and time \cite{Klein_1926}. This Lorentz covariant equation did not immediately become popular in the years it was introduced, since it had two fundamental flaws: negative energy
solutions and negative probabilities.  The invention of the Dirac equation in 1928 \cite{Dirac_1928}, and later its field theoretical interpretation \cite{Dirac_1930, Pauli_et_al_1934}, made the two flaws seen on the KG equation physically meaningful. Consequently, the KG equation emerged to a well-known tool in order to describe the relativistic spin-0 particle's dynamics.

KG equation was basically constructed via the linear momentum and rest mass quantities. The trivial solution to a free KG equation is just a plane wave. On the other hand in Nature, all particles interact, which necessitated the extension of the free KG equation to the realistic case. This can be obtained by the addition of a four-vector potential to the linear momentum and/or  coupling of or adding a scalar potential to the rest mass term \cite{Greiner_1990}. Note that a vector potential contains a time component in addition to its spatial components. Recently, L\"{u}tf\"{u}o\u{g}lu \textit{et al.} examined the scattering solutions of the KG equation in the existence of the coupling vector and scalar potential coupling
\cite{Lutfuoglu_et_al_2018}. There, the authors took the spatial component of the four vector to be zero while the time component to be non-zero in $(1+1)$ space-time. In a particular case, the magnitude of the non-zero component of the vector potential can be taken to be equal to the magnitude of the scalar potential. In literature, this case is called the spin-symmetry limit. In the pseudo-spin symmetry limit, unlike  the spin-symmetry limit, while the magnitudes of the two potentials  are equal to each other their relative signs  differ.

In many articles, with the use of different potential energies, the analytical and approximate solutions of the KG equation were investigated. Xie and Jia examined the KG equation's solution in higher spatial dimensions, with the Morse potential energy \cite{Xie_et_al_2015}. Yi \textit{et al.} obtained the bound state solution of the KG equation with equal scalar and vector form of the Rosen-Morse type potentials \cite{Yi_et_al_2004}. Soylu \textit{et al.} used the asymptotic iteration method to extend the solutions to the parity-time  symmetric version of the same  potential families \citep{Soylu_et_al_2008}. Zhang used an approximate method, namely the functional analysis method, to solve KG equation with equal magnitude scalar and vector Eckart potentials \cite{Zhang_2008}.  Saad \textit{et al.} studied the KG equation in the presence of scalar and vector potential in arbitrary dimensions and showed that if the both potentials were Coulombic, irrespective of whether they were equal to each other equal or
not, the analytic solutions could be obtained \cite{Saad_et_al_2008}. Momtazi \textit{et al.} used the Laplace transform approach and investigated the exact solution of the KG equation with unequal vector and scalar and Coulomb-like potentials \cite{Momtazi_et_al_2014}. Ikot \textit{et al.}
examined the bound state solutions of the KG equation under the Hylleraas Potential using the Nikiforov-Uvarov method \cite{Ikot_et_al_2012} and later investigated the modified Hylleraas potential with position dependent mass in D-dimensions \cite{Ikot_et_al_2013}. In 2014, Ikot \textit{et al.}  obtained the scattering and the bound state solutions of the one-dimensional KG equation with equal vector and scalar q-parameter hyperbolic P\"oschl-Teller potential energy \cite{Ikot_et_al_2015}. One year later, again Ikot \textit{et al.} used the factorization methods and supersymmetry to investigate the bound and scattering state of KG equation in D-dimensions with a deformed Hulth$\grave{e}$n plus deformed hyperbolical potential energy \cite{Ikot_et_al_2016}. Two of the authors of this paper,  Ikot and  L\"{u}tf\"{u}o\u{g}lu,  examined the KG equation solutions with an exponential-type molecule potential and discussed the thermodynamic properties in D dimensions \cite{Ikot_et_al_BCL_2016}. Das and Arda obtained the exact solutions of the KG equation for a charged particle under a spatially varying electromagnetic field \cite{Das_et_al_2017}. Very recently, Kisoglu and Sogut investigated the motion of a KG particle within an external
electromagnetic field when magnetic field was constant while the electric field depended on time \cite{Kisoglu_et_al_2018}.

Apart from the mentioned potential energies,   Woods-Saxon potential (WSP) energy possesses a significant attention in describing the Laws of Nature. Initially, it was put forward to calculate the differential cross section of the elastic scattering of protons by medium or heavy nuclei in 1954 \cite{Woods_et_al_1954}. Hou \textit{et al.} examined the bound state solutions  in the KG equation under the presence of vector and scalar WSP's \cite{Hou_et_al_1999}. Furthermore, Rojas and Villalba investigated the scattering solution of a KG particle without using the scalar coupling of the WSP \cite{Rojas_et_al_2005}.  Later, Hassanabadi \textit{et al.} include the scalar WSP to the vector potential and announced the scattering case results \cite{Hassanabadi_et_al_2013}. Satchler, in his book in the twelfth chapter proposed an extension to the WSP, namely Generalized Woods-Saxon Potential (GWSP), with taking account the surface interactions \cite{Satchler_1983}. Bayrak and Sahin used the GSWP in three spatial dimension in the KG equation and analyzed the bound state solutions \cite{Bayrak_et_al_2015}.  One of the authors of this paper, L\"{u}tf\"{u}o\u{g}lu, with his collaborators,  investigated the scattering case solution of the KG equation under the GWSP within spin and pseudospin symmetry limits \cite{Lutfuoglu_et_al_2018}. Then, he proved that in the pseudospin limit a bound state solution of the KG equation with the GWSP cannot be obtained \cite{Lutfuoglu_et_al_2018}. Furthermore, he used the obtained results in the nonrelativistic solutions of the GWSP \cite{Lutfuoglu_et_al_2016}  and compared the changes within the thermodynamic functions  in point of view in statistical mechanics \cite{Lutfuoglu_cjp_2018}.

Rosen and Morse introduced the Rosen-Morse potential(RMP) energy to investigate the vibrational states of polyatomic molecules such as ammonia molecule \cite{Rosen_et_al_1932}.   Since then, it is been widely employed on the molecular physics problems, i.e. \cite{Royappa_et_al_2006, Wang_et_al_2012}. Its improved version (IRMP), is used to calculate the  vibrational energies for the Cs dimer molecule and Na dimer molecule  \cite{Chen_et_al_2013, Hu_et_al_2014}. Very recently Jia \textit{et al.}  successfully predicted the molar entropy  and enthalpy values with the  Gibbs free energies for the nitrogen monoxide and gaseous phosphorus dimer in a wide temperature range by adopting the IRMP energy in their articles \cite{Jia_et_al2018a, Jia_et_al2018b, Peng_et_al_2018}.

In 2012,  Zhang et al. declared that they obtained a closed relation in between the IRMP and GWSP energies \cite{Zhang_et_al_2012}.  More precisely, in their study, they proved that the GWSP energy is identical to the IRMP energy  for diatomic molecules. This very important result yields that the GWSP energy can be used in real problems in molecular physics, too.

On the other hand, Ovando \textit{et al.}  recently investigated the equivalence of radial multi-parameter potential models for diatomic molecules \cite{Ovanda_et_al_2016}. They gave a class of multi-parameter exponential type potentials and showed  that different quantum interaction models used in describing diatomic molecules are the special cases of their proposed potential. Note that, Rafi \textit{et al.}  have shown there is no single potential function that can be regarded to be the best for all molecular states \cite{Rafi_et_al_2006}. They concluded that the more parameters in a potential function yield to the least percentage error in the molecular state. Furthermore, Jia \textit{et al.}  used a four-parameter diatomic molecular potential function instead of a three-parameter one and concluded that a four-parameter fit would be more accurate than a three-parameter one \cite{Jia_et_al_2017}.

The main motivation of the present work is to obtain an analytical solution for the KG equation by using  a multi-parameter q-Deformed Woods-Saxon type of potential energy. The proposed multi-parameterized potential energy can have many applications in nuclear and molecular physics as reported by Sovkov \textit{et al.} in \cite{Sovkov_et_al_2013}. Therefore, it is a good candidate for being used to explore the diatomic and polyatomic molecules  structures in molecular physics as a toy model.

Our manuscript is organized as follows. In sec. \ref{sec2:model}, we introduce the proposed potential energy and the KG equation with equal scalar and vector potentials.  In sec. \ref{sec3:generalsolution}, we solve the KG equation and obtain a general solution in terms of hypergeometric functions.  In sec. \ref{sec4:scattering}, we use the asymptotic behavior of the wave function and present the scattering state solution by taking account of the continuity conditions. Furthermore, we obtain a closed form expression of the probabilities of transmission and reflection and analyze the correlations with the potential parameters. In sec. \ref{sec5:boundstate}, we  derive the bound state solutions briefly. We assign arbitrary values to the parameters and we obtain an energy spectrum of a confined KG particle. We conclude the paper in sec. \ref{sec6:conclusion}.

\section{Model}\label{sec2:model}
In this article, we investigate the continuum and bound state solutions of the KG equation in the presence of the coupling of a scalar potential energy, $V_s$, and a vector potential energy, $V_v$, with nearly equal magnitudes, $V_v(x)=gV_s(x)\equiv V(x)$, within the strong regime, $g\approx1$,  in one spatial dimension
\begin{eqnarray}
  \frac{d^2\phi(x)}{dx^2}+\frac{1}{\hbar^2c^2}\Big[\big(E^2-M^2c^4\big)-2\big(E+Mc^2\big)V(x) \Big]\phi(x)&=& 0  \label{bett2.11}
\end{eqnarray}
Here $M$ is the rest mass of the particle, $c$ is the speed of light in vacuum and $\hbar$ is the Planck constant. The potential energy under the investigation has $26$ parameters and is given in the form of
\begin{eqnarray}
  V(x) &=& \theta(-x)\Bigg[\tilde{V}_0-\frac{\tilde{V}_1}{\tilde{q}+\tilde{p}e^{\tilde{\alpha}(x+\tilde{L})}}+ \frac{\tilde{V}_2}{\big(\tilde{q}+\tilde{p}e^{\tilde{\alpha}(x+\tilde{L})}\big)^2}
  +\tilde{\xi} \Bigg( \frac{\tilde{A}+\tilde{B}e^{\tilde{\alpha}(x+\tilde{L})}}{\tilde{q}+\tilde{p}
  e^{\tilde{\alpha}(x+\tilde{L})}}\Bigg)\nonumber \\
  &+& \tilde{\eta} \Bigg(\frac{\tilde{C}+\tilde{D}e^{\tilde{\alpha}(x+\tilde{L})}}
  {\tilde{q}+\tilde{p}e^{\tilde{\alpha}(x+\tilde{L})}}\Bigg)^2\Bigg]
  +\theta(x)\Bigg[V_0-\frac{V_1}{q+pe^{-\alpha(x-L)}}+ \frac{V_2}{\big(q+pe^{-\alpha(x-L)}\big)^2}\nonumber \\
  &+&\xi \Bigg( \frac{A+Be^{-\alpha(x-L)}}{q+pe^{-\alpha(x-L)}}\Bigg)+ \eta \Bigg( \frac{C+De^{-\alpha(x-L)}}{q+pe^{-\alpha(x-L)}}\Bigg)^2\Bigg]. \label{potenergy}
\end{eqnarray}
Here $\theta(\mp)$ represents the Heaviside step function. Note that the potential energy is not effective locally. In order to have a finite distance effective potential energy, which is inspired from the Woods-Saxon potential energy, the parameters $\tilde{V}_0$ and $V_0$ have to satisfy the conditions given by
\begin{eqnarray}
  \tilde{V}_0&=&\frac{1}{\tilde{q}}\big(\tilde{V}_1-\tilde{\xi} \tilde{A}\big)-\frac{1}{\tilde{q}^2}\big(\tilde{V}_2+\tilde{\eta} \tilde{C}^2\big), \label{Vzerotilde}\\
  V_0&=&\frac{1}{q}\big(V_1-\xi A\big)-\frac{1}{q^2}\big(V_2+\eta C^2\big)\label{Vzero}.
\end{eqnarray}

The potential energy can be set to be symmetric  via $x=0$ point. Consequently, the number of parameters needed to describe the system is halved. In the local effective case, $12$ independent parameters are left to be set. Among them, $V_1$, $V_2$, $A$, $B$, $C^2$, $D^2$ have dimensions of energy, while $q$,$p$, $\xi$, $\eta$ are dimensionless numbers. Note that, $\alpha$, is the measure of the slope of the potential well or barrier with dimension $m^{-1}$ while the final parameter, $L$, adjusts the effective length of the potential energy and has dimension of $m$.

\section{The general solution}\label{sec3:generalsolution}
Whether the potential energy is symmetric or not, the KG equation in both negative and positive regions are qualitatively the same. Therefore, we examine a solution in one region and generalize it to the other region. For $x<0$ case, we insert the locally effective potential energy  given in  Eq.~(\ref{potenergy}) into Eq.~(\ref{bett2.11}) and find
\begin{eqnarray}
  \frac{d^2\phi_L(x)}{dx^2}+\frac{\big(E^2-M^2c^4\big)}{\hbar^2c^2}\phi_L(x)-\frac{2\big(E+Mc^2\big)}{\hbar^2c^2}\Bigg[\tilde{V}_0-\frac{\tilde{V}_1}{\tilde{q}+\tilde{p}e^{\tilde{\alpha}(x+\tilde{L})}}+ \frac{\tilde{V}_2}{\big(\tilde{q}+\tilde{p}e^{\tilde{\alpha}(x+\tilde{L})}\big)^2}
  &&\nonumber \\+\tilde{\xi} \Bigg( \frac{\tilde{A}+\tilde{B}e^{\tilde{\alpha}(x+\tilde{L})}}{\tilde{q}+\tilde{p}
  e^{\tilde{\alpha}(x+\tilde{L})}}\Bigg)+ \tilde{\eta} \Bigg(\frac{\tilde{C}+\tilde{D}e^{\tilde{\alpha}(x+\tilde{L})}}
  {\tilde{q}+\tilde{p}e^{\tilde{\alpha}(x+\tilde{L})}}\Bigg)^2 \Bigg]\phi_L(x)= 0&.&\,\,\,  \label{bett3.1}
\end{eqnarray}
We introduce a transformation to the variable $x$ of the form $z=-\frac{\tilde{p}}{\tilde{q}}e^{\tilde{\alpha}(x+\tilde{L})}$, and we express the transformed wave function with $F(z)$. Then, we find  Eq.~(\ref{bett3.1}) is turned out into a dimensionless equation
\begin{eqnarray}
  \Bigg[\frac{d^2}{dz^2}+ \frac{1}{z} \frac{d}{dz}+ \frac{1}{z^{2}(1-z)^{2}}\big(\tilde{\omega}_0^2+\tilde{\omega}_1^2 z+\tilde{\omega}_2^2 z^2\big)\Bigg]F(z)&=&0. \label{bett3.2}
\end{eqnarray}
where
\begin{eqnarray}
\tilde{\omega}_0^2 &\equiv&  \frac{E^2-M^2c^4}{\tilde{\alpha}^2 \hbar^2c^2}, \label{bett3.3a}\\
\tilde{\omega}_1^2 &\equiv&   -\frac{2(E^2-M^2c^4)}{\tilde{\alpha}^2 \hbar^2c^2 }+ \frac{2(E+Mc^2)}{\tilde{\alpha}^2\hbar^2c^2 }
\Bigg[\frac{1}{\tilde{q}}\Big(\tilde{V}_1-\frac{2\tilde{V}_2}{\tilde{q}}\Big)-\tilde{\xi}\Big(\frac{\tilde{A}}
{\tilde{q}}-\frac{ \tilde{B}}{\tilde{p}}\Big)  - 2\tilde{\eta} \frac{\tilde{C}}{\tilde{q}} \Big(\frac{\tilde{C}}{\tilde{q}}-\frac{\tilde{D}}{\tilde{p}}\Big)\Bigg],\label{bett3.3b}\\
  \tilde{\omega}_2^2 &\equiv& \frac{(E^2-M^2c^4)}{\tilde{\alpha}^2 \hbar^2c^2 }- \frac{2(E+Mc^2)}{\tilde{\alpha}^2 \hbar^2c^2}
  \Bigg[\frac{1}{\tilde{q}}\Big(\tilde{V}_1-\frac{\tilde{V}_2}{\tilde{q}}\Big)-\tilde{\xi}\Big(\frac{\tilde{A}}
  {\tilde{q}}-\frac{ \tilde{B}}{\tilde{p}}\Big) - \tilde{\eta} \Big(\frac{\tilde{C}^2}{\tilde{q}^2}-\frac{\tilde{D}^2}{\tilde{p}^2}\Big)\Bigg].\label{bett3.3c}
\end{eqnarray}
At this point, we make an Ansatz. We take the wave function solution of Eq.~(\ref{bett3.2}) in the form,
\begin{eqnarray}
  F(z) &\equiv& z^{\tilde{\mu}} (1-z)^{\tilde{\nu}} \chi(z) \label{bett3.4}
\end{eqnarray}
Inserting Eq.~(\ref{bett3.4}) into Eq.~(\ref{bett3.2}), we obtain the hypergeometric equation
\begin{eqnarray}
  z(1-z)\chi''+ \Big[(1+2\tilde{\mu})-z(1+2\tilde{\mu}+2\tilde{\nu})\Big]\chi'-(\tilde{\mu}+\tilde{\nu}+\tilde{\lambda})
  (\tilde{\mu}+\tilde{\nu}-\tilde{\lambda})\chi &=& 0.  \label{bett3.5} \,\,\,\,\,\,\,\,\,\,\,\,
\end{eqnarray}
where the coefficients are found to be
\begin{eqnarray}
    \tilde{\mu}    &=& \mp i \tilde{\omega}_0 \\
                    &=& i \sqrt{\frac{E^2-M^2c^4}{\tilde{\alpha}^2\hbar^2c^2}}, \label{bett3.6a}\\
                    &\equiv& i \tilde{k}, \\
    \tilde{\nu}    &=& \frac{1}{2}\mp \sqrt{\frac{1}{4}+\tilde{\omega}_0^2+\tilde{\omega}_1^2+\tilde{\omega}_2^2 }  \\
                    &=& \frac{1}{2}\mp \sqrt{\frac{1}{4}+\frac{2(E+Mc^2)}{\tilde{\alpha}^2\hbar^2c^2 } \Bigg[\frac{1}{\tilde{q}^2}\bigg(\tilde{V}_2+\tilde{\eta} \tilde{C}^2\bigg)-\frac{\tilde{\eta} \tilde{D}}{\tilde{p}}\bigg(\frac{2\tilde{C}}{\tilde{q}}-\frac{\tilde{D}}{\tilde{p}}\bigg) \Bigg]},\label{bett3.6b} \\
    \tilde{\lambda}&=& i \tilde{\omega}_2 \\
    &=& i\sqrt{\frac{E^2-M^2c^4}{\tilde{\alpha}^2\hbar^2c^2}- \frac{2(E+Mc^2)}{\tilde{\alpha}^2\hbar^2c^2\tilde{p}^2}\Big(\tilde{\xi} \tilde{B}\tilde{p} + \tilde{\eta} \tilde{D}^2\Big)} \label{bett3.6c}
\end{eqnarray}
The solution of Eq.~(\ref{bett3.5}) is the hypergeometric function \cite{RefGradshytenRyzhikBook},
\begin{eqnarray}
\chi(z) &=& A_1 \,\,_2F_1[\tilde{\mu}+\tilde{\nu}+\tilde{\lambda},\tilde{\mu}+\tilde{\nu}-\tilde{\lambda},1+2\tilde{\mu};z] \nonumber \\&+& B_1 z^{-2\tilde{\mu}}\,\, _2F_1[-\tilde{\mu}+\tilde{\nu}+\tilde{\lambda},-\tilde{\mu}+\tilde{\nu}-\tilde{\lambda},1-2\tilde{\mu};z].  \label{bett3.7}
\end{eqnarray}
Therefore, the general solution in the negative region is found to be
\begin{eqnarray}
  F(z)&=& A_1 z^{\tilde{\mu}} (1-z)^{\tilde{\nu}} \,\,_2F_1[\tilde{\mu}+\tilde{\nu}+\tilde{\lambda},\tilde{\mu}+\tilde{\nu}-\tilde{\lambda},
  1+2\tilde{\mu};z] \nonumber \\
    &+& B_1  z^{-\tilde{\mu}} (1-z)^{\tilde{\nu}} \,\, _2F_1[-\tilde{\mu}+\tilde{\nu}+\tilde{\lambda},-\tilde{\mu}+\tilde{\nu}-\tilde{\lambda},1-2\tilde{\mu};z]. \label{bett3.8}
\end{eqnarray}
The general solution in positive region, $\phi_R(x)$, can be written directly by using  Eq.~(\ref{bett3.8}) as
\begin{eqnarray}
  G(y)&=& C_1 y^{\mu} (1-z)^{\nu} \,\,_2F_1[\mu+ \nu+\lambda, \mu+ \nu-\lambda, 1+2 \mu;y] \nonumber \\
    &+& D_1  z^{-\mu} (1-z)^{\nu} \,\, _2F_1[-\mu+\nu+\lambda,-\mu+\nu-\lambda,1-2\mu;y]. \label{bett3.9}
\end{eqnarray}
Note that, in the positive region the coordinate transformation $y=-\frac{p}{q}e^{-\alpha(x-L)}$ and  the modified wave function $\phi_R(x)\rightarrow G(y)$ are used. Moreover, the parameters $\mu$, $\nu$ and $\lambda$ are defined by
\begin{eqnarray}
    \mu  &=& i \sqrt{\frac{E^2-M^2c^4}{\alpha^2 \hbar^2c^2}}, \label{bett3.10a}\\
    &\equiv& ik, \\
    \nu    &=&  \frac{1}{2}\mp \sqrt{\frac{1}{4}+\frac{2(E+Mc^2)}{\alpha^2\hbar^2c^2} \Bigg[\frac{1}{q^2}\bigg(V_2+\eta C^2\bigg)-\frac{\eta D}{p}\bigg(\frac{2C}{q}-\frac{D}{p}\bigg) \Bigg]},\label{bett3.10.} \\
    \lambda &=&  i\sqrt{\frac{E^2-M^2c^4}{\alpha^2 \hbar^2c^2}- \frac{2(E+Mc^2)}{\alpha^2 \hbar^2c^2 p^2}\Big(\xi B p + \eta D^2\Big)} \label{bett3.10c}
\end{eqnarray}
\section{Continuum State Solution}\label{sec4:scattering}

In this manuscript, we assumed that an incident particle approached from negative infinity. Therefore,  reflected and transmitted wave functions should be found. The asymptotic behavior at $x\rightarrow -\infty$ is obtained from Eq.~(\ref{bett3.8}) such that
\begin{eqnarray}
  \phi_L(x\rightarrow -\infty)  &\approx& A_1 \bigg(-\frac{\tilde{p}}{\tilde{q}}\bigg)^{i\tilde{k}} e^{i\tilde{k}\tilde{\alpha}(x+\tilde{L})}+B_1 \bigg(-\frac{\tilde{p}}{\tilde{q}}\bigg)^{-i\tilde{k}} e^{-i\tilde{k}\tilde{\alpha}(x+\tilde{L})}. \label{bett4.1}
\end{eqnarray}
On the other hand at the positive infinity, the asymptotic behaviour is written by using  Eq.~(\ref{bett3.9})
\begin{eqnarray}
\phi_R(x\rightarrow \infty) &\approx& C_1 \bigg(-\frac{p}{q}\bigg)^{ik} e^{-ik\alpha(x-L)}+D_1 \bigg(-\frac{p}{q}\bigg)^{-ik} e^{ik\alpha(x-L)}. \label{bett4.2}
\end{eqnarray}
As a consequence of the determination of the direction of the approach of the particle to the potential energy barrier of well, at negative infinity, only transmitted  wave functions could be seen. Therefore, $C_1=0$.

\subsection{Continuity conditions in symmetric potential energy}
From now on we only investigate the potential energy with its symmetric form. Therefore,
\begin{eqnarray}
  \tilde{\mu} &=& \mu, \\
  \tilde{\nu} &=& \nu, \\
  \tilde{\lambda} &=& \lambda.
\end{eqnarray}

In order to have  self consistent results, the wave function must be well defined and continuous in every point. Therefore, we have to examine the continuity conditions
\begin{eqnarray}
  \phi_L(x)\bigg|_{x\rightarrow 0^-} &=& \phi_R(x)\bigg|_{x\rightarrow 0^+},  \label{bett4.3}\\
  \frac{d\phi_L(x)}{dx}\bigg|_{x\rightarrow 0^-} &=& \frac{d\phi_R(x)}{dx}\bigg|_{x\rightarrow 0^+}. \label{bett4.4}
\end{eqnarray}
Note that, these conditions can be expressed with their transformed forms, too.
\begin{eqnarray}
   F(z)\bigg|_{z=z_0} &=& G(y)\bigg|_{y=y_0},  \label{bett4.5}\\
   \bigg(z\frac{dF(z)}{dz}\bigg|_{z=z_0} &=& -\bigg(y\frac{dG(y)}{dy}\bigg|_{y=y_0}, \label{bett4.6}
\end{eqnarray}
while $z_0=-\frac{\tilde{p}}{\tilde{q}} e^{\tilde{\alpha} \tilde{L}}=y_0=-\frac{p}{q} e^{\alpha L}\equiv t_0$.

If we use the first condition, namely  Eq.~(\ref{bett4.5}), we find
\begin{eqnarray}
  \frac{D_1}{ A_1}-\frac{B_1}{A_1}&=& (t_0)^{2\mu}\frac{M_1}{M_2} \label{bett4.7}
\end{eqnarray}
where
\begin{eqnarray}
  M_1 &\equiv&  \,\,_2F_1[\mu+\nu+\lambda,\mu+\nu-\lambda,1+2\mu; t_0]  \label{bett4.8}\\
  M_2 &\equiv& \,\, _2F_1[-\mu+\nu+\lambda,-\mu+\nu-\lambda,1-2\mu; t_0]\label{bett4.9}
\end{eqnarray}
We employ the property of the hypergeometric function \cite{RefGradshytenRyzhikBook},
\begin{eqnarray}
 _2F_1[a,b,c;x]&=&\frac{\Gamma(c)\Gamma(b-a)}{\Gamma(b)\Gamma(c-a)}(-x)^{-a}  \,\,_2F_1\Big[a,1+a-c,1+a-b;\frac{1}{x}\Big]\nonumber \\
 &+& \frac{\Gamma(c)\Gamma(a-b)}{\Gamma(a)\Gamma(c-b)}(-x)^{-b}
  \,\,_2F_1\Big[b,1+b-c,1+b-a;\frac{1}{x}\Big]. \,\, \label{bett4.10}
\end{eqnarray}
and we obtain $M_1$
\begin{eqnarray}
  M_1 &=&  \frac{\Gamma(1+2\mu)\Gamma(-2\lambda)}{\Gamma(\mu+\nu-\lambda)\Gamma(1+\mu-\nu-\lambda)}
  (t_0)^{-\mu-\nu-\lambda}(-1)^{-\mu-\nu-\lambda}  \nonumber \\
  && \times \,\,_2F_1\Bigg[\mu+\nu+\lambda,-\mu+\nu+\lambda,1+2\lambda; \frac{1}{t_0}\Bigg]\nonumber \\
 &+& \frac{\Gamma(1+2\mu)\Gamma(2\lambda)}{\Gamma(\mu+\nu+\lambda)\Gamma(1+\mu-\nu+\lambda)}
 (t_0)^{-\mu-\nu+\lambda} (-1)^{-\mu-\nu+\lambda}\nonumber \\
  && \times
  \,\,_2F_1\Bigg[\mu+\nu-\lambda,-\mu+\nu-\lambda,1-2\lambda; \frac{1}{t_0}\Bigg]. \,\, \label{bett4.11}
\end{eqnarray}
and $M_2$
\begin{eqnarray}
  M_2 &=& \frac{\Gamma(1-2\mu)\Gamma(-2\lambda)}{\Gamma(-\mu+\nu-\lambda)\Gamma(1-\mu-\nu-\lambda)}
  \Big(t_0\Big)^{\mu-\nu-\lambda} (-1)^{\mu-\nu-\lambda} \nonumber \\
  && \times \,\,_2F_1\Bigg[-\mu+\nu+\lambda,\mu+\nu+\lambda,1+2\lambda; \frac{1}{t_0}\Bigg]\nonumber \\
 &+& \frac{\Gamma(1-2\mu)\Gamma(2\lambda)}{\Gamma(-\mu+\nu+\lambda)\Gamma(1-\mu-\nu+\lambda)}
 \Big(t_0\Big)^{\mu-\nu+\lambda} (-1)^{\mu-\nu+\lambda}\nonumber \\
  && \times
  \,\,_2F_1\Bigg[-\mu+\nu-\lambda,\mu+\nu-\lambda,1-2\lambda; \frac{1}{t_0}\Bigg]. \,\,  \label{bett4.12}
\end{eqnarray}
Here we assign new abbreviations as
\begin{eqnarray}
  N_1 &\equiv& \,\,_2F_1\Bigg[\mu+\nu+\lambda,-\mu+\nu+\lambda,1+2\lambda; \frac{1}{t_0}\Bigg], \label{bett4.13} \\
  N_2 &\equiv& \,\,_2F_1\Bigg[\mu+\nu-\lambda,-\mu+\nu-\lambda,1-2\lambda; \frac{1}{t_0}\Bigg],
  \label{bett4.14} \\
  N_3 &\equiv& \,\,_2F_1\Bigg[-\mu+\nu+\lambda,\mu+\nu+\lambda,1+2\lambda;\frac{1}{t_0}\Bigg],
  \label{bett4.15} \\
  N_4 &\equiv& \,\,_2F_1\Bigg[-\mu+\nu-\lambda,\mu+\nu-\lambda,1-2\lambda;\frac{1}{t_0}\Bigg],
  \label{bett4.16}
\end{eqnarray}
and
\begin{eqnarray}
  S_1 &\equiv& \frac{\Gamma(1+2\mu)\Gamma(-2\lambda)}{\Gamma(\mu+\nu-\lambda)\Gamma(1+\mu-\nu-\lambda)}, \label{bett4.17}\\
  S_2 &\equiv& \frac{\Gamma(1+2\mu)\Gamma(2\lambda)}{\Gamma(\mu+\nu+\lambda)\Gamma(1+\mu-\nu+\lambda)},
  \label{bett4.18} \\
  S_3 &\equiv& \frac{\Gamma(1-2\mu)\Gamma(-2\lambda)}{\Gamma(-\mu+\nu-\lambda)\Gamma(1-\mu-\nu-\lambda)},
  \label{bett4.19} \\
  S_4 &\equiv& \frac{\Gamma(1-2\mu)\Gamma(2\lambda)}{\Gamma(-\mu+\nu+\lambda)\Gamma(1-\mu-\nu+\lambda)}.
  \label{bett4.20}
\end{eqnarray}
By using the new abbreviations, we re-script $M_1$ and $M_2$
\begin{eqnarray}
  M_1 &=&  S_1 N_1 (-1)^{-\mu-\nu-\lambda} (t_0)^{-\mu-\nu-\lambda}+ S_2 N_2 (-1)^{-\mu-\nu+\lambda} (t_0)^{-\mu-\nu+\lambda},  \label{M1}\\
  M_2 &=& S_3 N_3 (-1)^{\mu-\nu-\lambda}  (t_0)^{\mu-\nu-\lambda} +S_4 N_4 (-1)^{\mu-\nu+\lambda} (t_0)^{\mu-\nu+\lambda}.  \label{M2}
\end{eqnarray}
Finally, we revise the result that is given in  Eq.~(\ref{bett4.7}) as
\begin{eqnarray}
   \frac{B_1}{A_1} &=& \frac{D_1}{ A_1}- \frac{S_1 N_1 (-1)^{-\mu-\lambda} (t_0)^{-\lambda}+ S_2 N_2 (-1)^{-\mu+\lambda} (t_0)^{\lambda}}{S_3 N_3 (-1)^{\mu-\lambda}  (t_0)^{-\lambda} +S_4 N_4 (-1)^{\mu+\lambda} (t_0)^{\lambda}}  \label{bett4.21}.
\end{eqnarray}
To solve  the second condition given in Eq.~(\ref{bett4.6}), we need to use the relation of the derivative of the hypergeometric function \cite{RefGradshytenRyzhikBook}
\begin{eqnarray}
  \frac{_2F_1[a,b,c,t]}{dt} &=& \frac{ab}{c}  {}_2F_1[a+1,b+1,c+1,t]. \label{bett4.22}
\end{eqnarray}
We find
\begin{eqnarray}
    \frac{B_1}{A_1}=-\frac{D_1}{A_1}-(t_0)^{2\mu}  \frac{\Bigg[\bigg(\frac{\mu}{t_0} - \frac{\nu}{(1-t_0)} \bigg) M_1       + \frac{(\mu+\nu)^2-\lambda^2}{1+2\mu} M_3\Bigg]}{\Bigg[\bigg(-\frac{\mu}{ t_0} - \frac{\nu}{(1-t_0)} \bigg) M_2 +\frac{(-\mu+\nu)^2-\lambda^2}{1-2\mu}  M_4\Bigg]} \label{bett4.23}
\end{eqnarray}
where
\begin{eqnarray}
  M_3 &=&  \frac{\Gamma(2+2\mu)\Gamma(-2\lambda)}{\Gamma(1+\mu+\nu-\lambda) \Gamma(1+\mu-\nu-\lambda)}
  (t_0)^{-1-\mu-\nu-\lambda}(-1)^{-1-\mu-\nu-\lambda}  \nonumber \\
  && \times \,\,_2F_1\Bigg[1+\mu+\nu+\lambda,-\mu+\nu+\lambda,1+2\lambda; \frac{1}{t_0}\Bigg]\nonumber \\
 &+& \frac{\Gamma(2+2\mu)\Gamma(2\lambda)}{\Gamma(1+\mu+\nu+\lambda)\Gamma(1+\mu-\nu+\lambda)}
 ( t_0)^{-1-\mu-\nu+\lambda} (-1)^{-1-\mu-\nu+\lambda}\nonumber \\
  && \times
  \,\,_2F_1\Bigg[1+\mu+\nu-\lambda,-\mu+\nu-\lambda,1-2\lambda; \frac{1}{t_0}\Bigg]. \,\, \label{bett4.24}
\end{eqnarray}
and
\begin{eqnarray}
  M_4 &=& \frac{\Gamma(2-2\mu)\Gamma(-2\lambda)}{\Gamma(1-\mu+\nu-\lambda)\Gamma(1-\mu-\nu-\lambda)}
  (t_0)^{-1+\mu-\nu-\lambda}(-1)^{-1+\mu-\nu-\lambda}  \nonumber \\
  && \times \,\,_2F_1\Bigg[1-\mu+\nu+\lambda,\mu+\nu+\lambda,1+2\lambda; \frac{1}{t_0}\Bigg]\nonumber \\
 &+& \frac{\Gamma(2-2\mu)\Gamma(2\lambda)}{\Gamma(1-\mu+\nu+\lambda)\Gamma(1-\mu-\nu+\lambda)}
 (t_0)^{-1+\mu-\nu+\lambda} (-1)^{-1+\mu-\nu+\lambda}\nonumber \\
  && \times
  \,\,_2F_1\Bigg[1-\mu+\nu-\lambda,\mu+\nu-\lambda,1-2\lambda; \frac{1}{t_0}\Bigg]. \,\, \label{bett4.25}
\end{eqnarray}
We define the following abbreviation to obtain a simpler expression as we have done above.
\begin{eqnarray}
  N_5 &\equiv& \,\,_2F_1\Bigg[1+\mu+\nu+\lambda,-\mu+\nu+\lambda,1+2\lambda; \frac{1}{t_0}\Bigg], \label{bett4.26} \\
  N_6 &\equiv& \,\,_2F_1\Bigg[1+\mu+\nu-\lambda,-\mu+\nu-\lambda,1-2\lambda; \frac{1}{t_0}\Bigg], \label{bett4.27} \\
  N_7 &\equiv& \,\,_2F_1\Bigg[1-\mu+\nu+\lambda,\mu+\nu+\lambda,1+2\lambda; \frac{1}{t_0}\Bigg],  \label{bett4.28} \\
  N_8 &\equiv& \,\,_2F_1\Bigg[1-\mu+\nu-\lambda,\mu+\nu-\lambda,1-2\lambda; \frac{1}{t_0}\Bigg].  \label{bett4.29}
\end{eqnarray}
and
\begin{eqnarray}
  S_5 &\equiv& \frac{\Gamma(2+2\mu)\Gamma(-2\lambda)}{\Gamma(1+\mu+\nu-\lambda) \Gamma(1+\mu-\nu-\lambda)},
  \label{bett4.30} \\
  S_6 &\equiv& \frac{\Gamma(2+2\mu)\Gamma(2\lambda)}{\Gamma(1+\mu+\nu+\lambda)\Gamma(1+\mu-\nu+\lambda)},
  \label{bett4.31} \\
  S_7 &\equiv& \frac{\Gamma(2-2\mu)\Gamma(-2\lambda)}{\Gamma(1-\mu+\nu-\lambda)\Gamma(1-\mu-\nu-\lambda)},
  \label{bett4.32} \\
  S_8 &\equiv& \frac{\Gamma(2-2\mu)\Gamma(2\lambda)}{\Gamma(1-\mu+\nu+\lambda)\Gamma(1-\mu-\nu+\lambda)}.
  \label{bett4.33}
\end{eqnarray}
We re-script $M_3$ and $M_4$ in terms of the new abbreviations.
\begin{eqnarray}
  M_3 &=&  S_5 N_5(-1)^{-1-\mu-\nu-\lambda} (t_0)^{-1-\mu-\nu-\lambda}+ S_6 N_6 (-1)^{-1-\mu-\nu+\lambda} (t_0)^{-1-\mu-\nu+\lambda}, \label{bett4.34} \\
  M_4 &=& S_7 N_7 (-1)^{-1+\mu-\nu-\lambda} (t_0)^{-1+\mu-\nu-\lambda}+ S_8 N_8 (-1)^{-1+\mu-\nu+\lambda}(t_0)^{-1+\mu-\nu+\lambda}. \label{bett4.35}
\end{eqnarray}
By adding and subtracting Eq.~(\ref{bett4.21}) and Eq.~(\ref{bett4.23}), we find
\begin{eqnarray}
    \frac{D_1}{ A_1}&=& \frac{(t_0)^{2\mu}}{2}\Bigg[\frac{M_1}{M_2}-\frac{\bigg(\frac{\mu}{t_0} - \frac{\nu}{(1-t_0)} \bigg) M_1       + \frac{(\mu+\nu)^2-\lambda^2}{1+2\mu} M_3}{\bigg(-\frac{\mu}{ t_0} - \frac{\nu}{(1-t_0)} \bigg) M_2 +\frac{(-\mu+\nu)^2-\lambda^2}{1-2\mu}  M_4}\Bigg], \label{bett4.36}\\
    \frac{B_1}{ A_1}&=& -\frac{(t_0)^{2\mu}}{2}\Bigg[\frac{M_1}{M_2}+\frac{\bigg(\frac{\mu}{t_0} - \frac{\nu}{(1-t_0)} \bigg) M_1       + \frac{(\mu+\nu)^2-\lambda^2}{1+2\mu} M_3}{\bigg(-\frac{\mu}{ t_0} - \frac{\nu}{(1-t_0)} \bigg) M_2 +\frac{(-\mu+\nu)^2-\lambda^2}{1-2\mu}  M_4}\Bigg]. \label{bett4.37}
\end{eqnarray}
The wave function can be expressed in terms of $A_1$ by inserting the solutions found in Eq.~(\ref{bett4.36}) and Eq.~(\ref{bett4.37}) into Eq.~(\ref{bett3.8}) and Eq.~(\ref{bett3.9}).

\subsection{Transmission and Reflection  probabilities and their dependence on the potential energy parameters}
The transmission probability, $T$, and the reflection probability, $R$, are defined with
\begin{eqnarray}
T\equiv  \frac{D_1}{A_1} \Bigg(\frac{D_1}{A_1}\Bigg)^*, \label{bett4.38} \\
R\equiv  \frac{B_1}{A_1} \Bigg(\frac{B_1}{A_1}\Bigg)^*. \label{bett4.39}
\end{eqnarray}
where their sum is a conserved quantity which equals to one.  In order to prove that our results satisfy this conservation law, we assign arbitrarily chosen positive values  to $12$ parameters as given in Table~{\ref{parameters}} with natural units where $\hbar=c=1$.

\begin{figure}[hb]
  \centering
  \includegraphics[width=\textwidth]{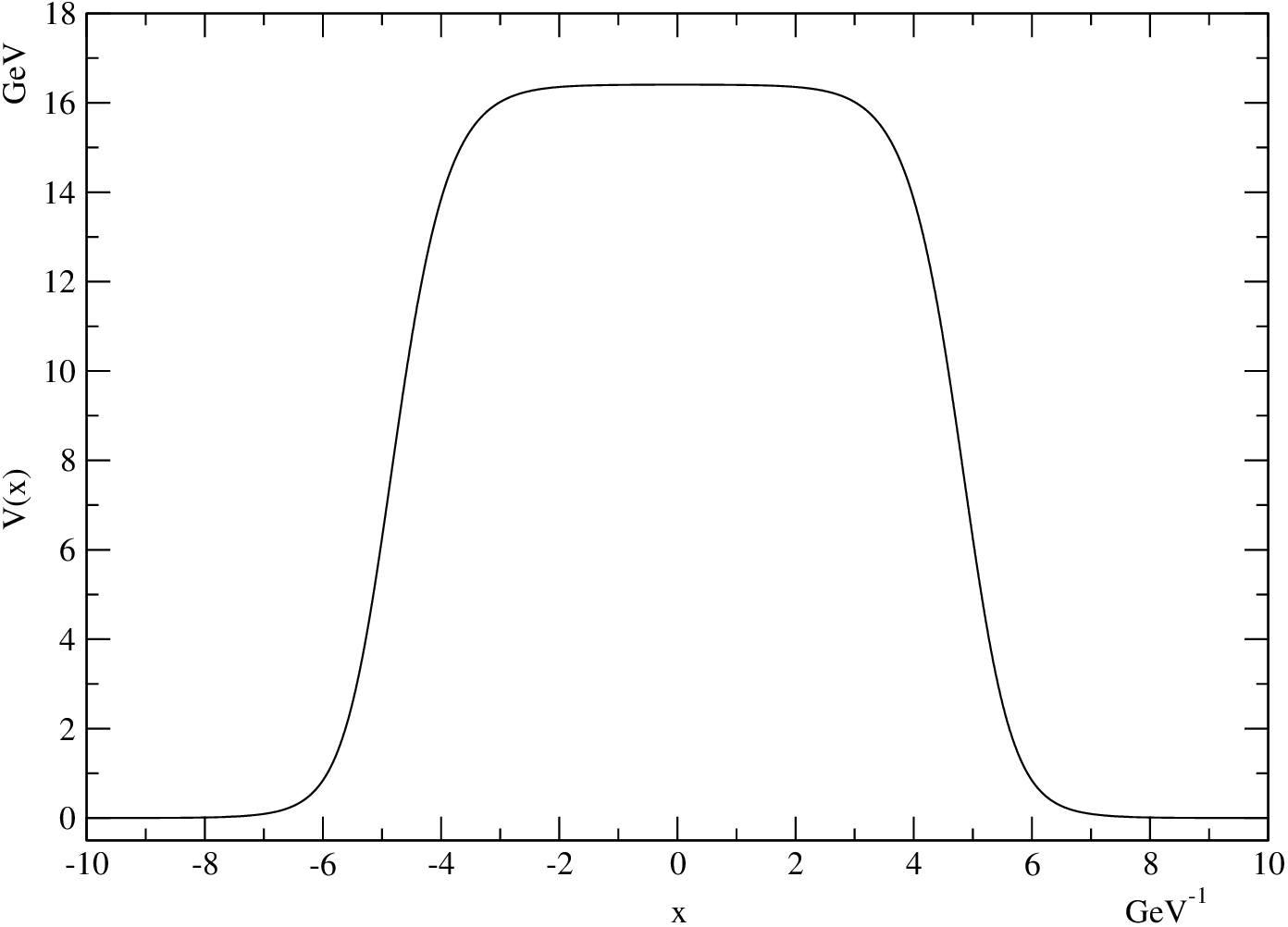}\\
  \caption{The investigated potential barrier that is obtained with the assigned parameters in Table~\ref{parameters}, versus the distance in the continuum case.}\label{Potentialwell}
\end{figure}

\begin{figure}
  \centering
  \includegraphics[width=\textwidth]{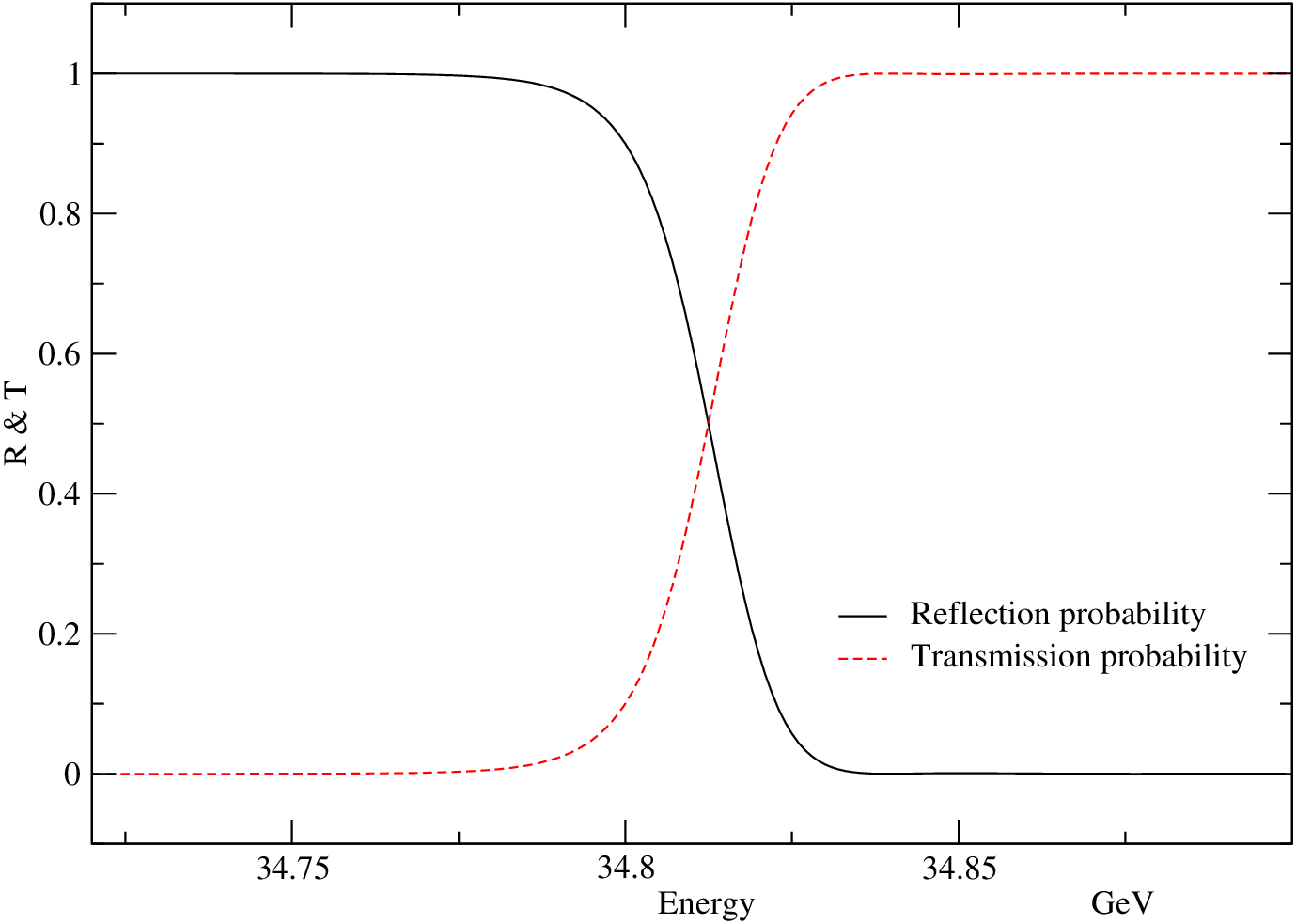}\\
  \caption{The transmission and reflection probabilities of the particle versus the scattered particle's energy.}\label{TransReflecversusEnergy}
\end{figure}

\begin{figure}[htp]
  \centering
  \includegraphics[width=\textwidth]{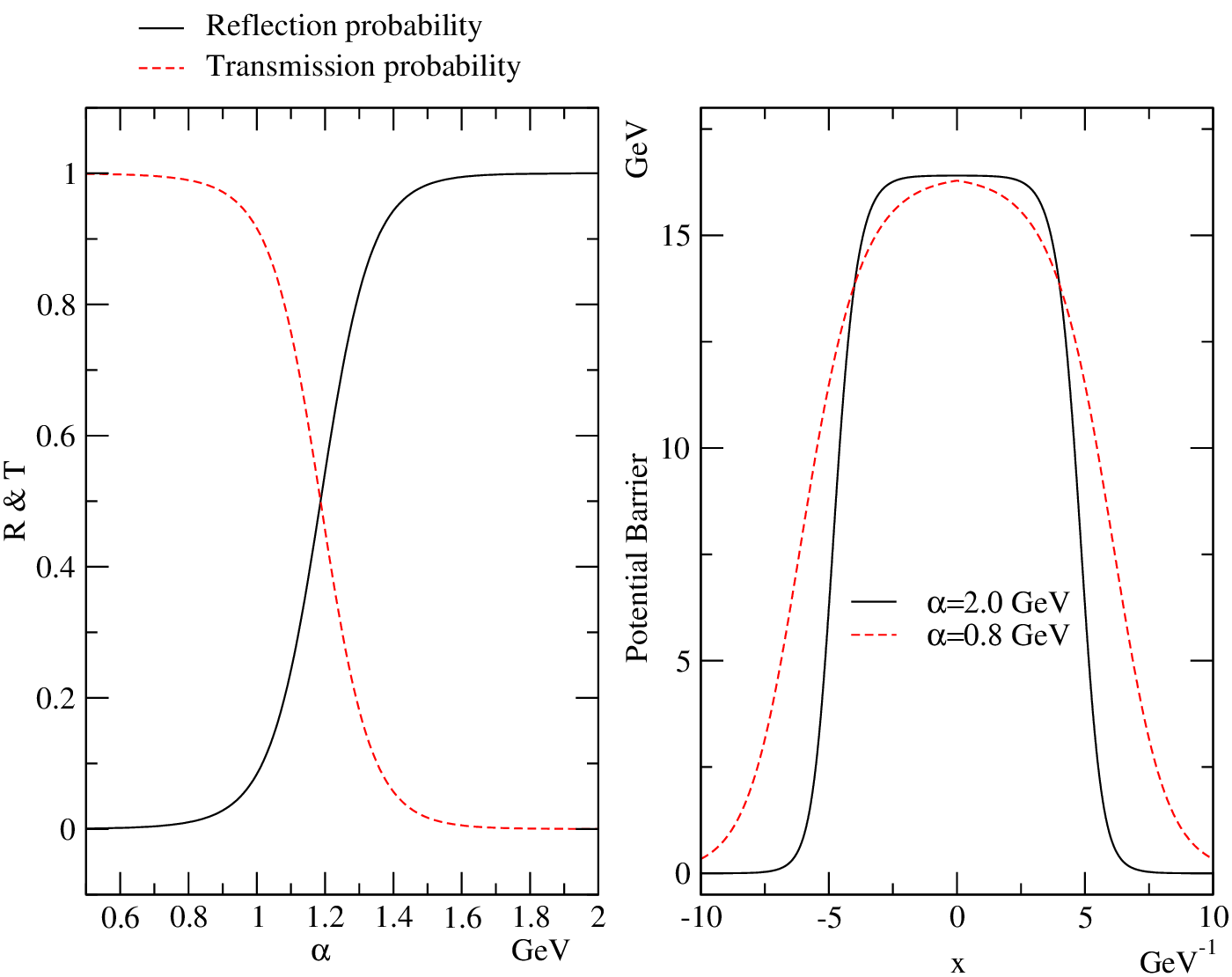}\\
  \caption{$\alpha$ parameter versus the transmission and reflection probabilities  of the incident particles with energy $E=34.75$ $GeV$.}\label{ModTransReflecCoefficientsversusALPHA}
\end{figure}

\begin{figure}[htp]
  \centering
  \includegraphics[width=\textwidth]{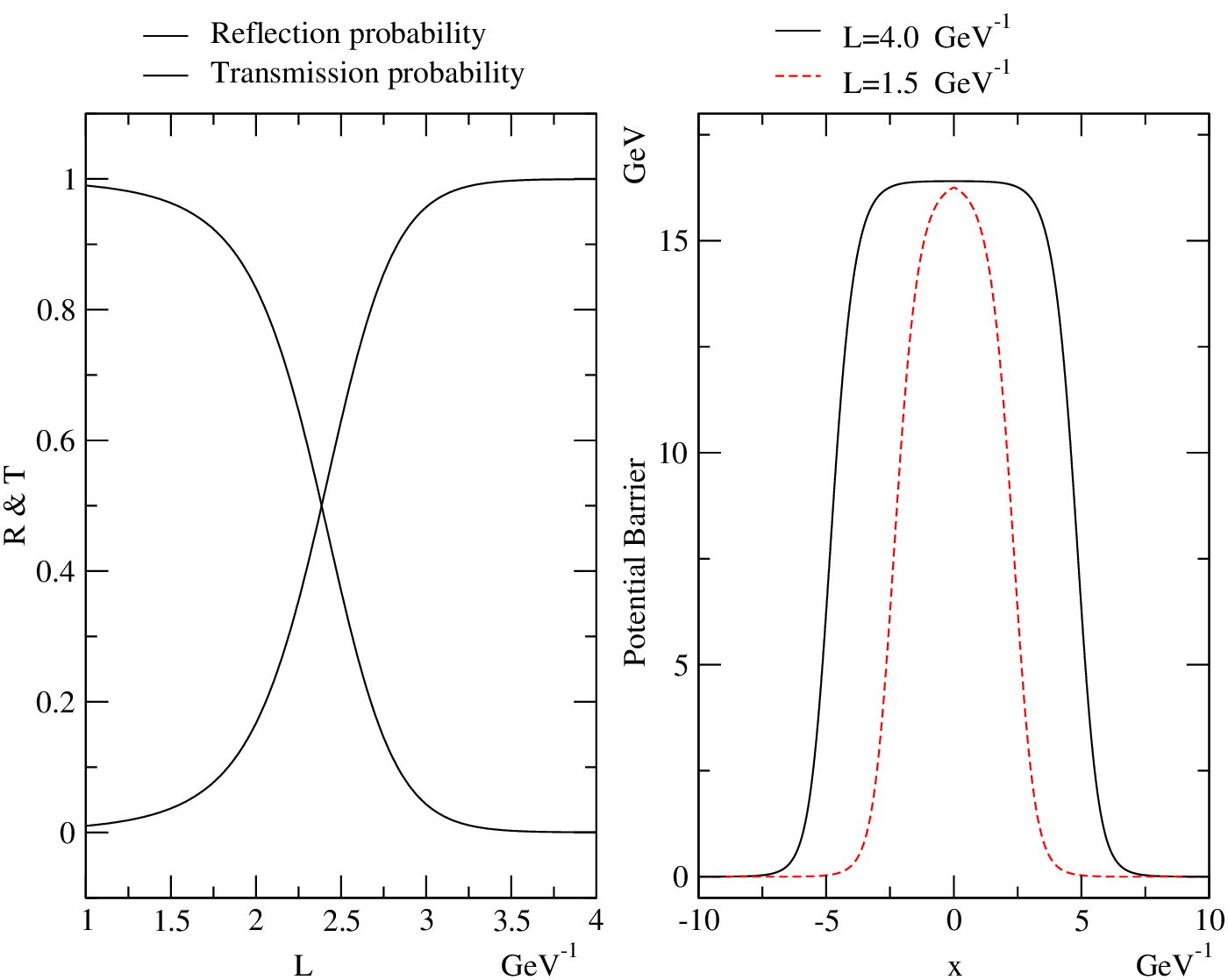}\\
  \caption{The effective distance parameter, $L$, versus the transmission and reflection probabilities  of the incident particles with energy $E=34.75$ $GeV$.}\label{ModTransReflecCoefficientsversusR}
\end{figure}

\begin{figure}[htp]
  \centering
  \includegraphics[width=\textwidth]{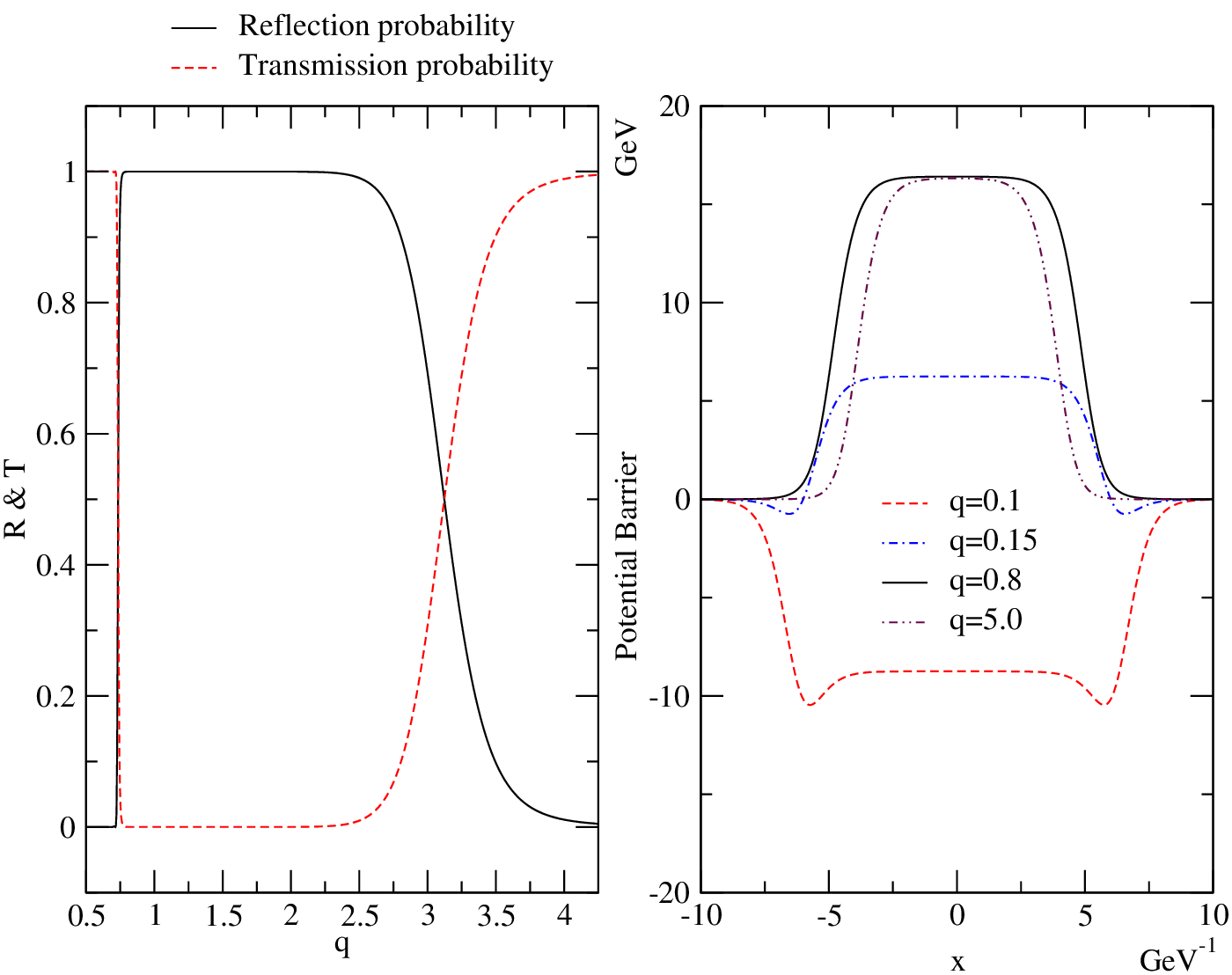}\\
  \caption{$q$ parameter versus the transmission and reflection probabilities  of the incident particles with energy $E=34.75$ $GeV$.}\label{ModTransReflecCoefficientsversusQ}
\end{figure}

\begin{figure}[htp]
  \centering
  \includegraphics[width=\textwidth]{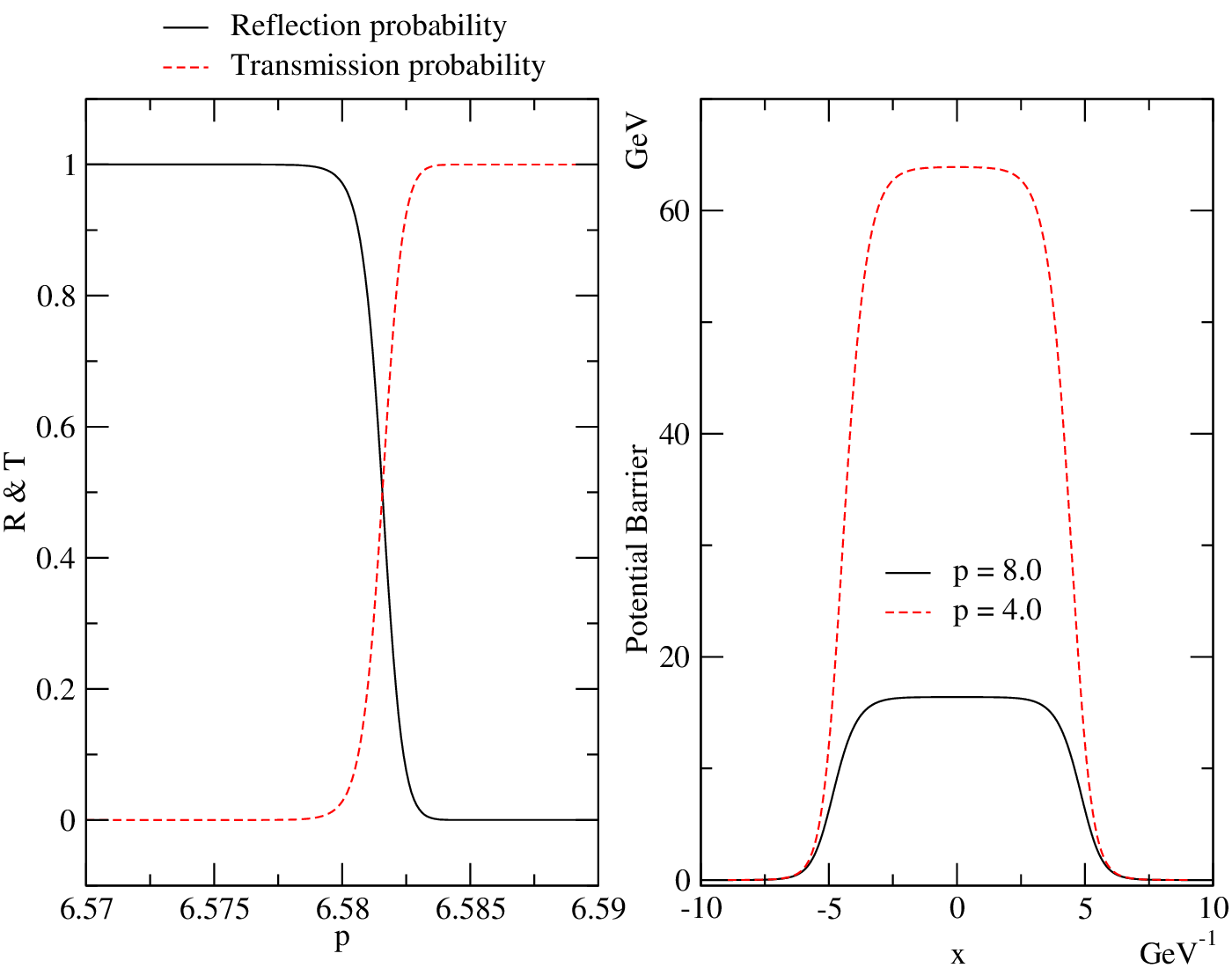}\\
  \caption{$q$ parameter versus the transmission and reflection probabilities  of the incident particles with energy $E=50.00$ $GeV$.}\label{ModTransReflecCoefficientsversusP}
\end{figure}

\begin{figure}[htp]
  \centering
  \includegraphics[width=\textwidth]{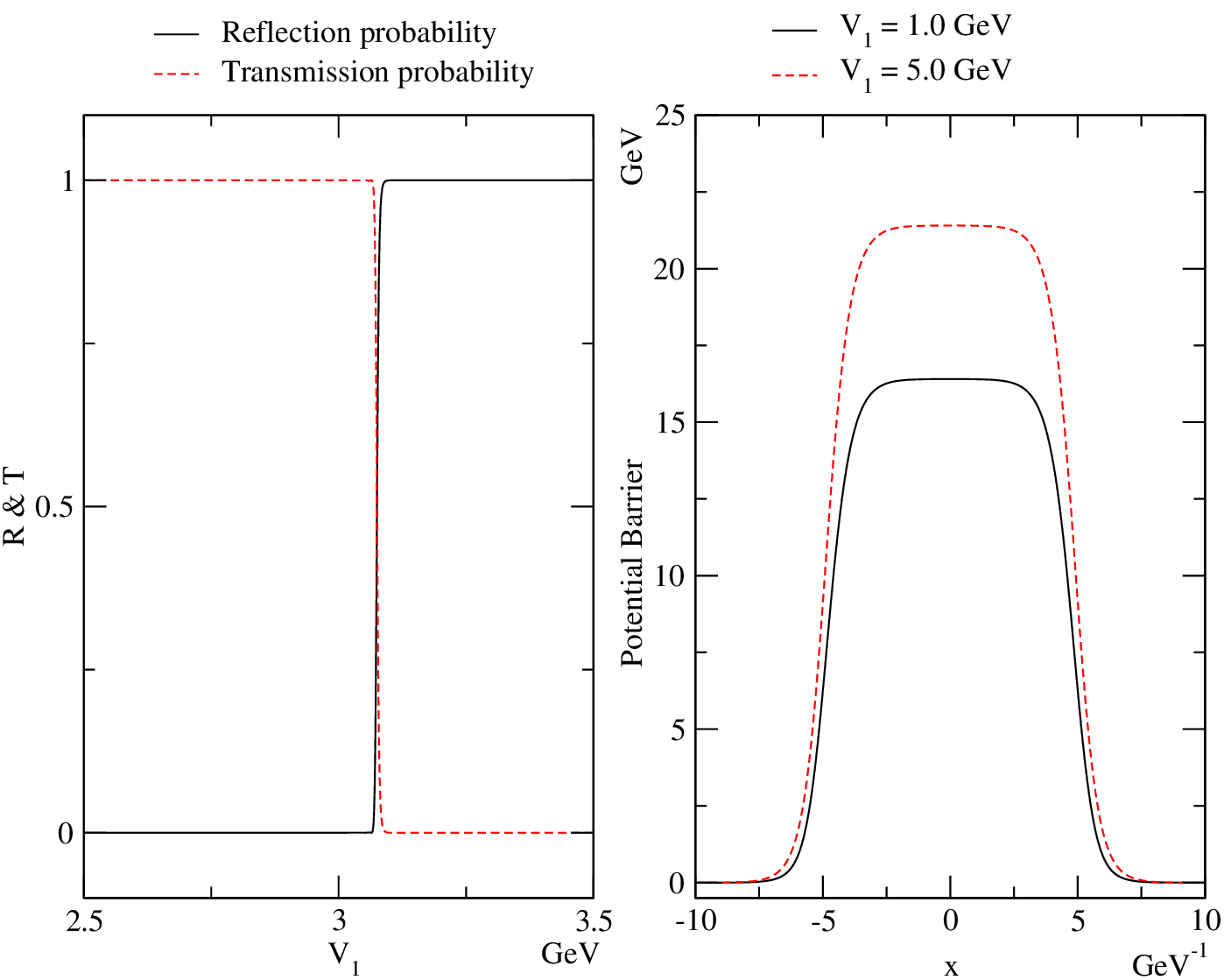}\\
  \caption{$V_1$ parameter versus the transmission and reflection probabilities  of the incident particles with energy $E=40.00$ $GeV$.}\label{ModTransReflecCoefficientsversusV1}
\end{figure}

\begin{figure}[htp]
  \centering
  \includegraphics[width=\textwidth]{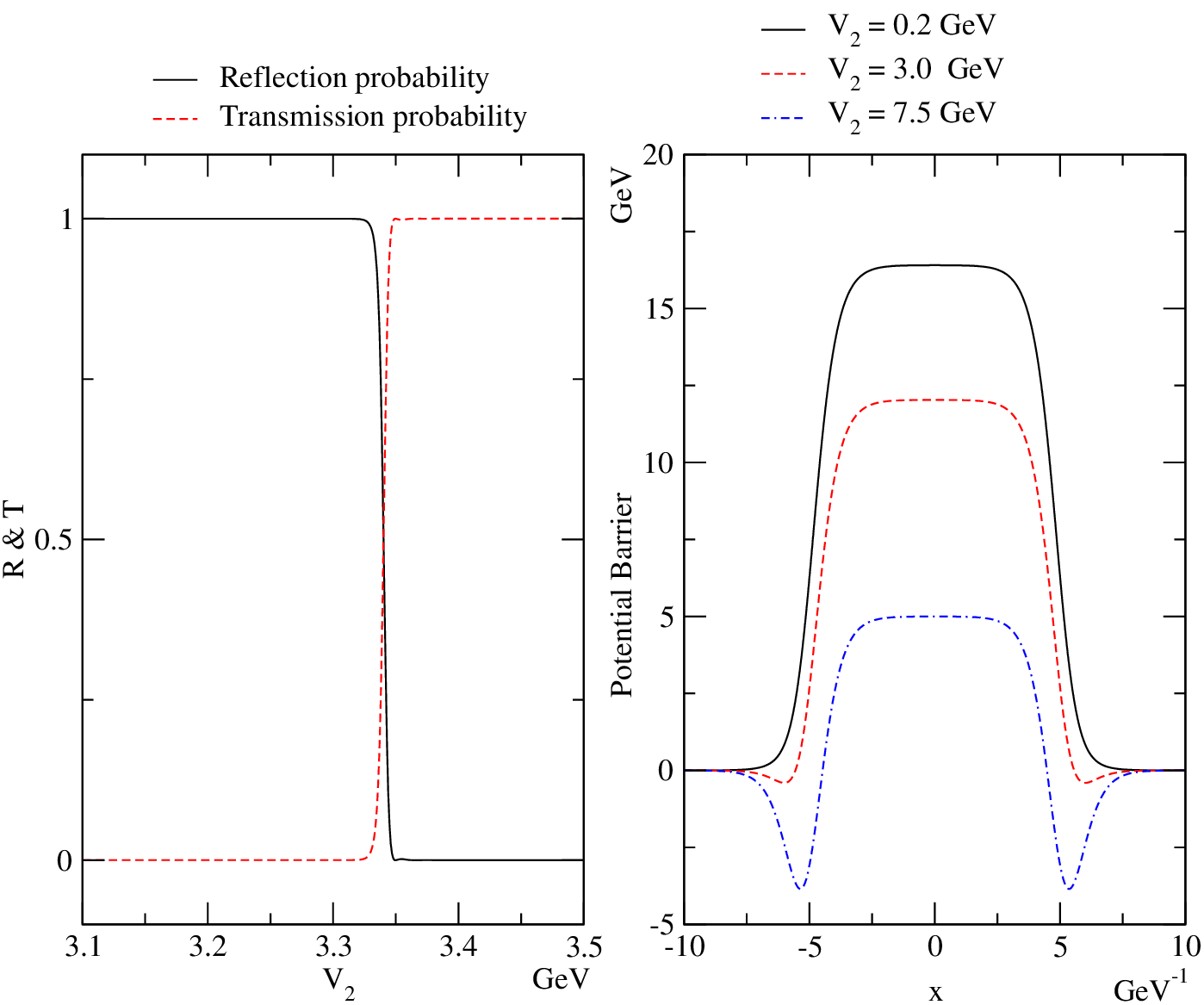}\\
  \caption{$V_2$ parameter versus the transmission and reflection probabilities  of the incident particles with energy $E=25.00$ $GeV$.}\label{ModTransReflecCoefficientsversusV2}
\end{figure}

\begin{figure}[htp]
  \centering
  \includegraphics[width=\textwidth]{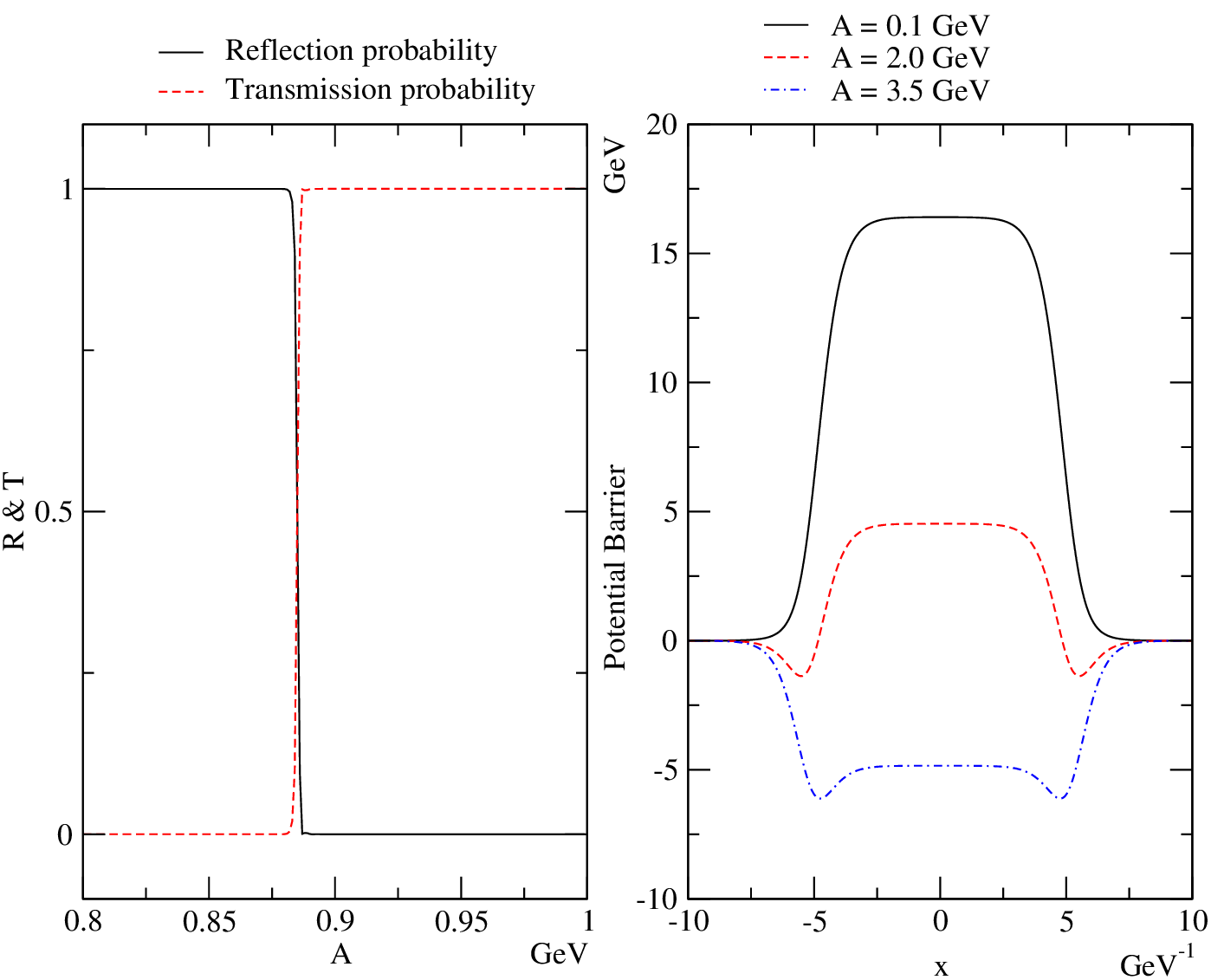}\\
  \caption{$A$ parameter versus the transmission and reflection probabilities  of the incident particles with energy $E=25.00$ $GeV$.}\label{ModTransReflecCoefficientsversusA}
\end{figure}

\begin{figure}[htp]
  \centering
  \includegraphics[width=\textwidth]{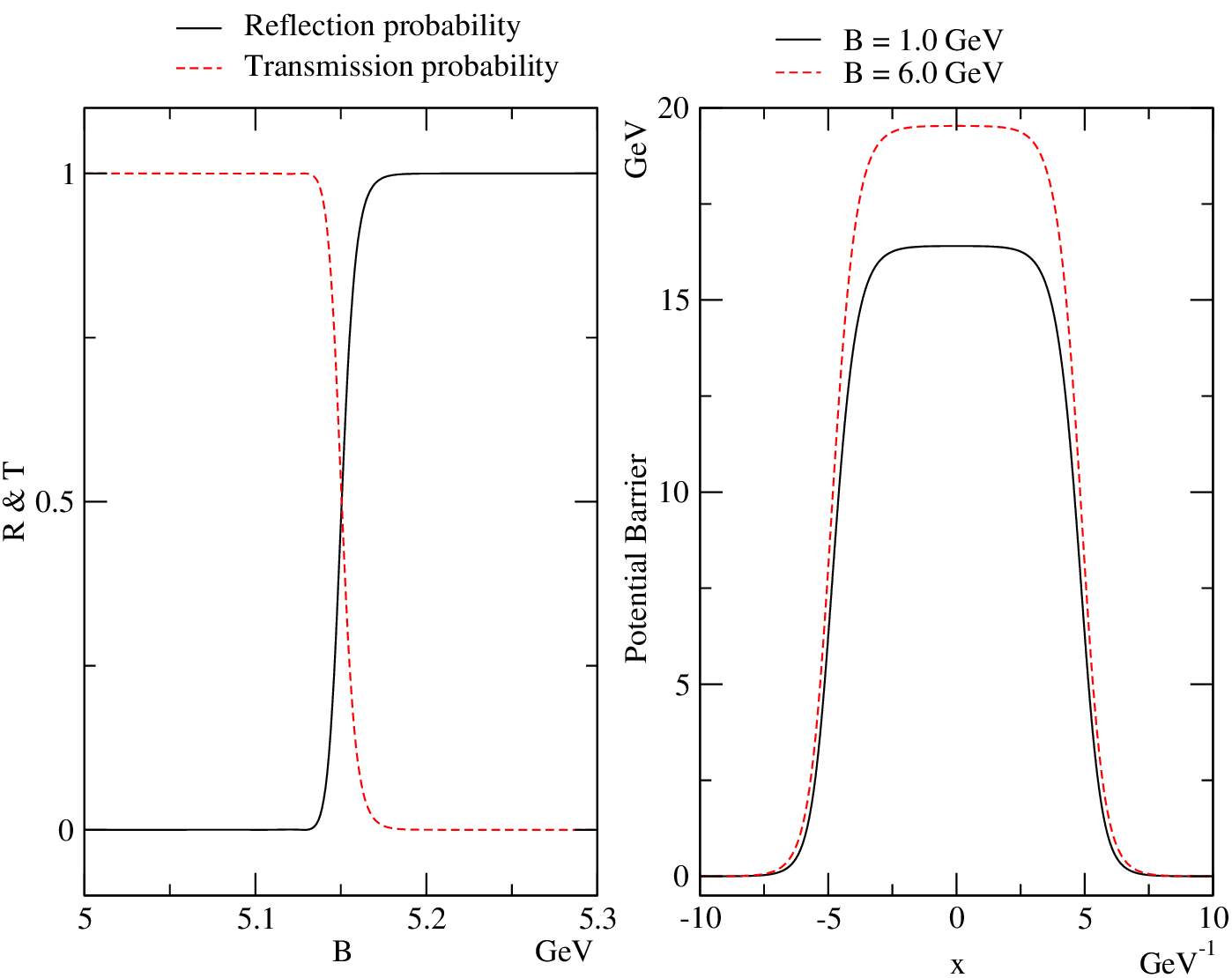}\\
  \caption{$B$ parameter versus the transmission and reflection probabilities  of the incident particles with energy $E=40.00$ $GeV$.}\label{ModTransReflecCoefficientsversusB}
\end{figure}

\begin{figure}[htp]
  \centering
  \includegraphics[width=\textwidth]{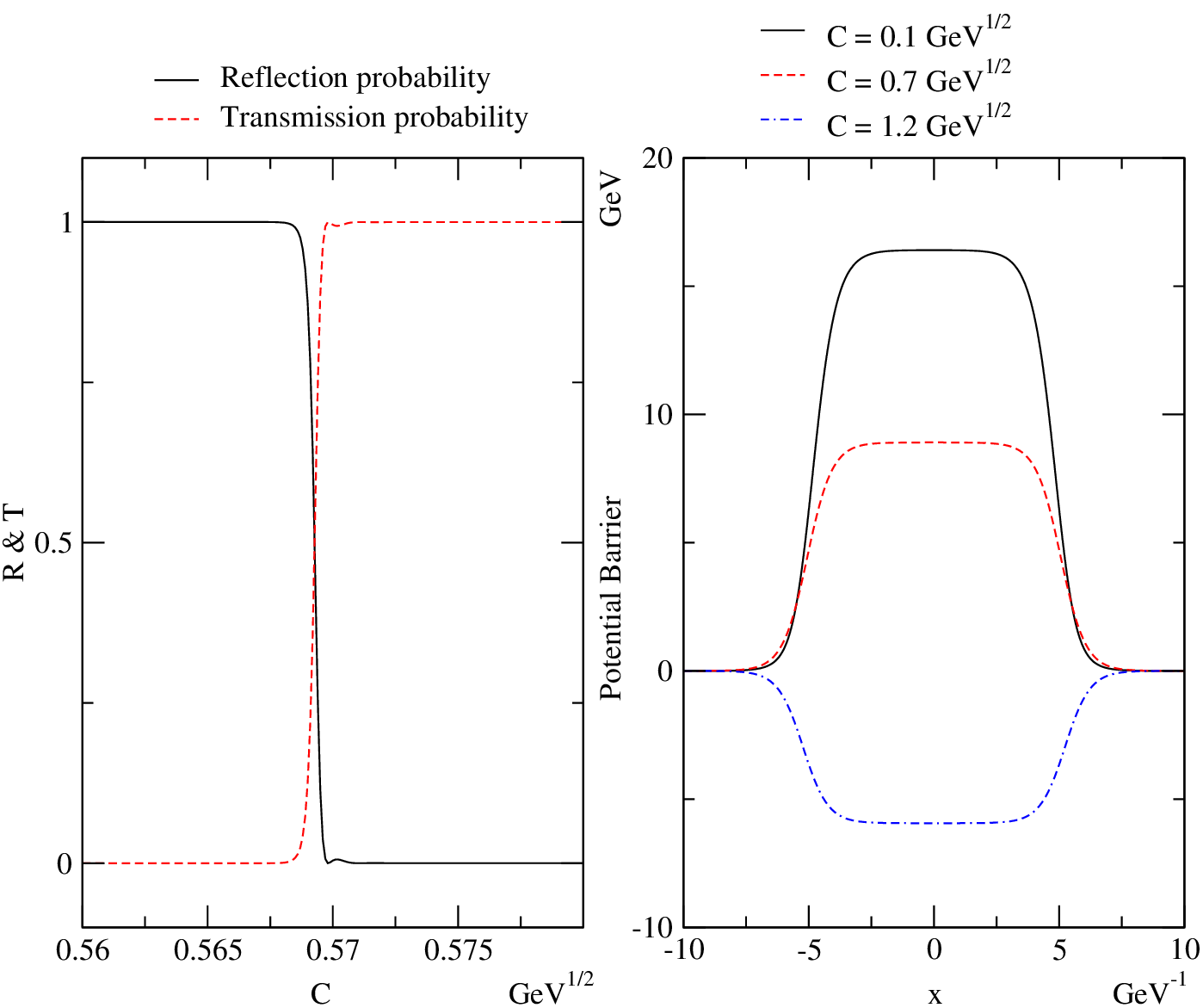}\\
  \caption{$V_2$ parameter versus the transmission and reflection probabilities  of the incident particles with energy $E=25.00$ $GeV$.}\label{ModTransReflecCoefficientsversusC}
\end{figure}

\begin{figure}[htp]
  \centering
  \includegraphics[width=\textwidth]{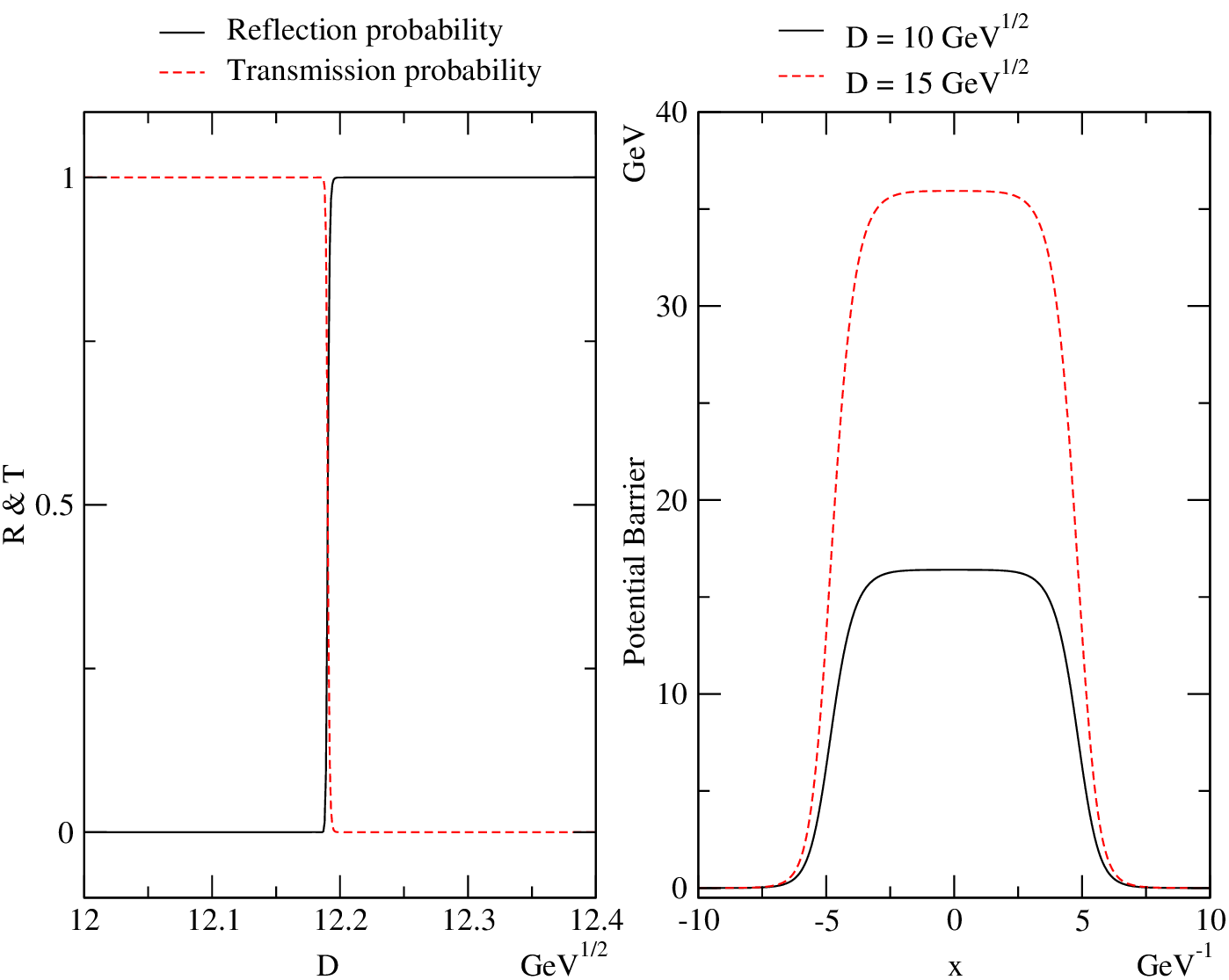}\\
  \caption{$V_2$ parameter versus the transmission and reflection probabilities  of the incident particles with energy $E=50.00$ $GeV$.}\label{ModTransReflecCoefficientsversusD}
\end{figure}

Eq.~(\ref{Vzerotilde}) or Eq.~(\ref{Vzero})  is used  to calculate the dependent parameter that makes the potential energy to approach to zero at infinities, and we find that $\tilde{V}_0 = V_0=\frac{5}{32}GeV$. We plot the investigated potential energy barrier in Fig.~\ref{Potentialwell} as a function of distance. Then, the transmission and reflection probabilities are calculated numerically for a scattering particle that has mass $M=2$ $GeV$. The variation of the probabilities versus the energy of the scattered particle is shown in Fig~{\ref{TransReflecversusEnergy}}. The condition of the conservation of the total probability is verified. In Fig~{\ref{ModTransReflecCoefficientsversusALPHA}} the variation of the probabilities via the $\alpha$ parameter is investigated. Note that the incident particle is assumed to have an energy $34.75$ $GeV$. Since the potential energy barrier mimics the WSP, the parameter $\alpha$  corresponds to the reciprocal diffusion parameter. We find that the decrease of the diffusion parameter, or the increase of $\alpha$, makes the potential barrier to become less penetrable as shown in the second column. Therefore, the transmission probability goes to zero while the reflection probability tends to one. The affect of the effective radius of the potential barrier on the transmission and reflection probability is examined in Fig~{\ref{ModTransReflecCoefficientsversusR}}. The increase of $L$ widens the potential barrier, as shown in the second column, which results in the decrease of the transmission probability, as one expects. The $q$ dependence of the probabilities is investigated in Fig~{\ref{ModTransReflecCoefficientsversusQ}}. When $q$ is less than a critical value, which is correlated with the other parameters, in this case among $0.1$ and $0.15$,  the  potential energy barrier turns to be a well as shown in the second column. A sufficiently energetic particle, in this case with energy $34.75$ $GeV$, has a unit transmission probability until $q=0.75$. With the increase of the $q$ parameter, the potential barrier shrinks. Therefore, after another critical value, the sufficiently energetic incident particle starts to have a non zero transmission probability. On the other hand, $p$ dependence of the transmission and reflection probabilities are observed in Fig~{\ref{ModTransReflecCoefficientsversusP}}. Contrarily to the $q$ parameter, $p$ parameter does not possess a critical value that transforms a characteristic change in the potential barrier. Therefore, depending of the incident particle energy, the reflection probability goes to zero after a value of $p$ parameter. The transmission and reflection probabilities versus the amplitude parameters, namely $V_1$, $V_2$, $A$, $B$, $C$ and $D$, are investigated in Fig~{\ref{ModTransReflecCoefficientsversusV1}}, Fig~{\ref{ModTransReflecCoefficientsversusV2}}, Fig~{\ref{ModTransReflecCoefficientsversusA}}, Fig~{\ref{ModTransReflecCoefficientsversusB}}, Fig~{\ref{ModTransReflecCoefficientsversusC}}, Fig~{\ref{ModTransReflecCoefficientsversusD}}, respectively. The change of the amplitude parameters are illustrated in the second columns. The probabilities versus the two other  parameters, $\xi$ and $\eta$, are not given  in the manuscript since they do not involve an extraordinary information.

\section{The Bound State Solution\label{sec5:boundstate}}

In the bound state case, unlike the scattering case, only one parameter is defined differently since the particle's energy spectrum occurs within the interval of $-Mc^2<E<Mc^2$.
\begin{eqnarray}
\tilde{\omega}_0^2 &\equiv&  -\frac{E^2-M^2c^4}{\tilde{\alpha}^2 \hbar^2c^2}.
\end{eqnarray}
Consequently, $\tilde{\mu}$ becomes a real number
\begin{eqnarray}
    \tilde{\mu} &=& \tilde{K}, \label{bett5.2}
\end{eqnarray}
since the definition of the wave number $\tilde{K}$ is
\begin{eqnarray}
  \tilde{K} &\equiv& \sqrt{-\frac{E^2-M^2c^4}{\tilde{\alpha}^2 \hbar^2 c^2}}. \label{bett5.1}
\end{eqnarray}
Note that, since it is the square root of a real number, it can be a positive or a negative real number. Here, we prefer to use the positive value of $\tilde{K}$.

Among the other unitless coefficients, $\tilde{\nu}$, given in Eq.~(\ref{bett3.6b}), can be a real or an imaginary number similar to the scattering case. On the other hand, $\tilde{\lambda}$, given in Eq.~(\ref{bett3.6c}), becomes a real number.

The particle's wave function has to decay to zero exponentially outside of the potential well. Therefore, the asymptotic behavior dictates that the coefficients $A_1$ and $C_1$ in the general solutions, given in Eq.~(\ref{bett3.8}) and Eq.~(\ref{bett3.9}), have to be zero. Consequently, the wave function owns a solution in negative and positive region as
\begin{eqnarray}
  F(z)&=& B_1  z^{-\tilde{\mu}} (1-z)^{\tilde{\nu}} \,\, _2F_1[-\tilde{\mu}+\tilde{\nu}+\tilde{\lambda},-\tilde{\mu}+\tilde{\nu}-\tilde{\lambda},1-2\tilde{\mu};z], \label{bett5.3} \\
  G(y)&=&  D_1  z^{-\mu} (1-z)^{\nu} \,\, _2F_1[-\mu+\nu+\lambda,-\mu+\nu-\lambda,1-2\mu;y]. \label{bett5.4}
\end{eqnarray}

By using the first continuity conditions we get
\begin{eqnarray}
 (B_1-D_1)t_0^{-\mu}(1-t_0)^{\nu}  M_2 &=& 0, \label{bett5.5}
\end{eqnarray}
and from the second condition we obtain
\begin{eqnarray}
 (B_1+D_1)t_0^{-\mu}(1-t_0)^{\nu} \Bigg[\bigg(-\frac{\mu}{t_0} - \frac{\nu}{(1-t_0)} \bigg) M_2
       + \frac{(-\mu+\nu)^2-\lambda^2}{1-2\mu} M_4\Bigg]  &=&0. \label{bett5.6}
\end{eqnarray}
Note that, $M_2$ and $M_4$ possess  the same definitions as given in  Eq.~(\ref{M2}) and Eq.~(\ref{bett4.35}).

\subsection{Energy Eigenvalues}
The symmetric structure of the potential energy allows the separation of the energy spectrum into two subsets which do not intersect. One the subsets, that is called even energy spectrum, is calculated by $B_1=D_1$. In this case, Eq.~(\ref{bett5.5}) becomes zero identically and Eq.~(\ref{bett5.6}) should be solved numerically. The corresponding wave function is obtained from the linear combination of Eq.~(\ref{bett5.3}) and Eq.~(\ref{bett5.4}). The second subset, namely odd energy spectrum, is obtained by anti-symmetrization of the normalization constant, $B_1=-D_1$. In this case, Eq.~(\ref{bett5.6}) is satisfied identically, while Eq.~(\ref{bett5.5}) need to be solved, numerically. Similarly to the even case, odd wave functions are derived from  Eq.~(\ref{bett5.3}) and Eq.~(\ref{bett5.4}) via their corresponding eigenvalue.

\subsection{Results and Discussions}

\begin{figure}[hb]
  \centering
  \includegraphics[width=\textwidth]{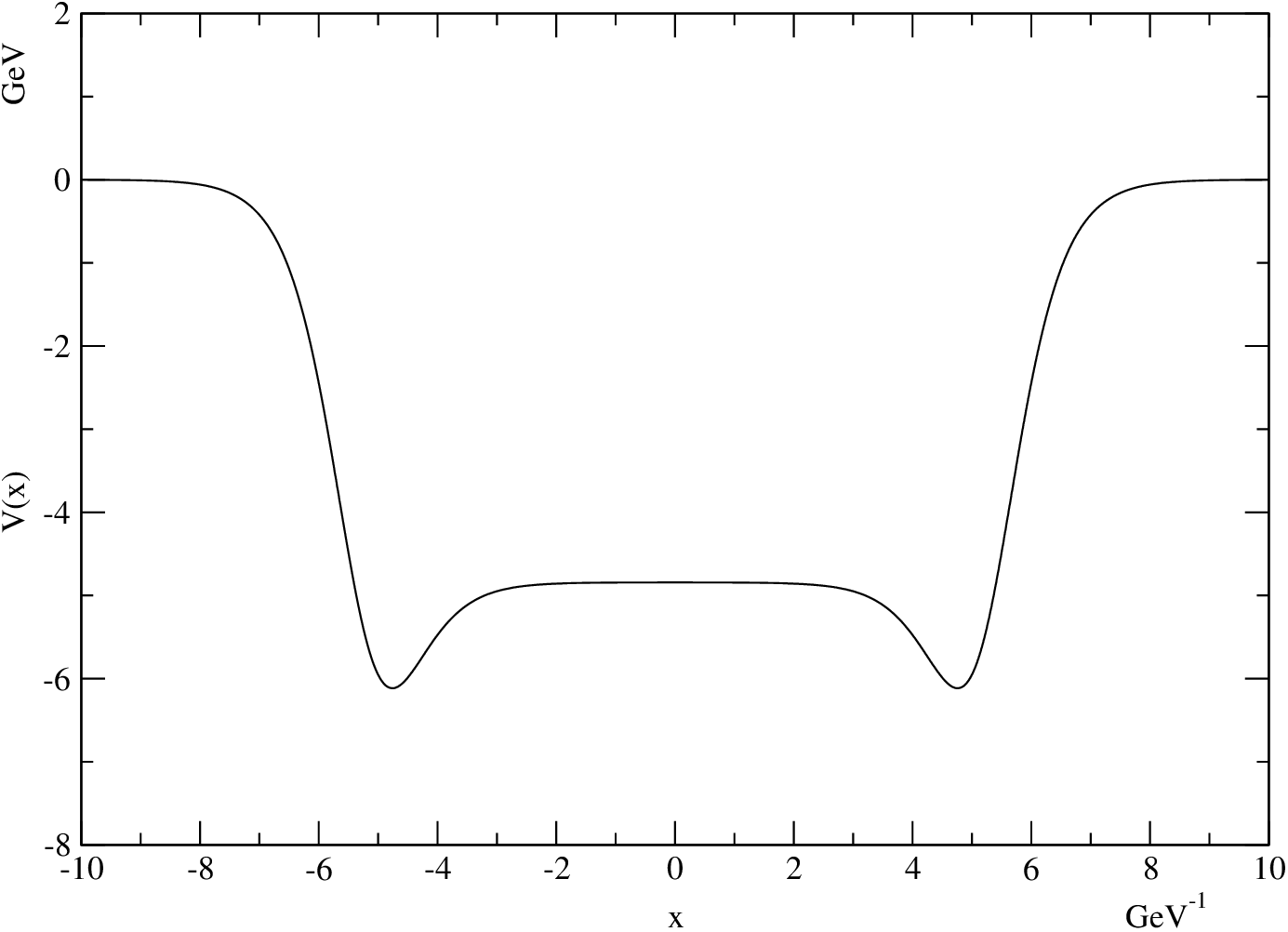}\\
  \caption{The potential well for the parameters given in the Table~{\ref{parameters}} with the exceptional value of the parameter $A=3.5$ $GeV$. }\label{BoundPotentialwell}
\end{figure}

In order to obtain a potential energy well, we use the same parameters given in  the Table~{\ref{parameters}} with only one exception. Namely, We choose $A$ parameter equal to $3.5$ $GeV$  instead of $0.1$ $GeV$.  In Fig.~\ref{BoundPotentialwell}, we illustrate the potential energy well.

\begin{figure}[hb]
  \centering
  \includegraphics[width=\textwidth]{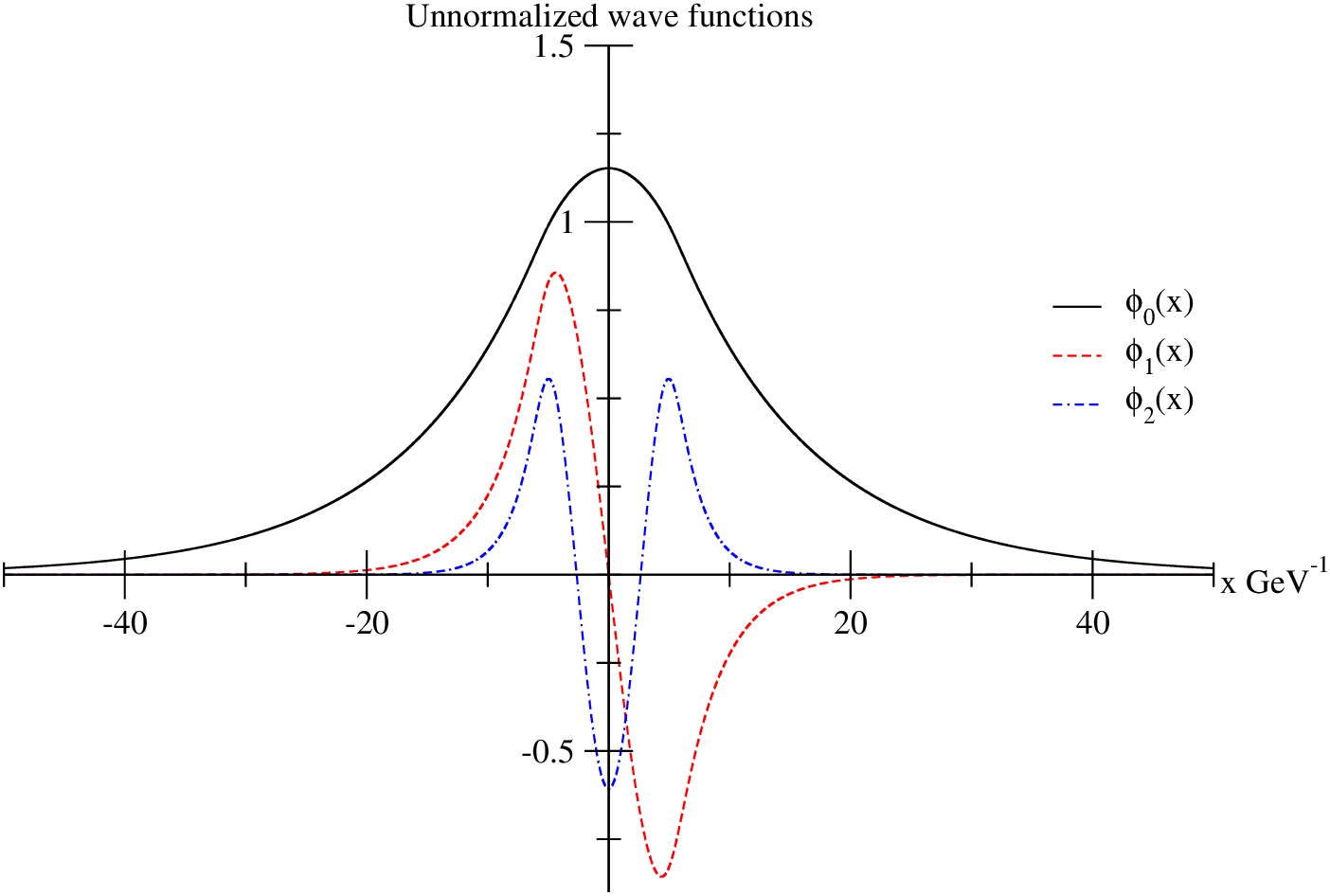}\\
  \caption{The first three unnormalized eigenfunctions.}\label{Boundstates-threewave}
\end{figure}

\begin{figure}[hb]
  \centering
  \includegraphics[width=\textwidth]{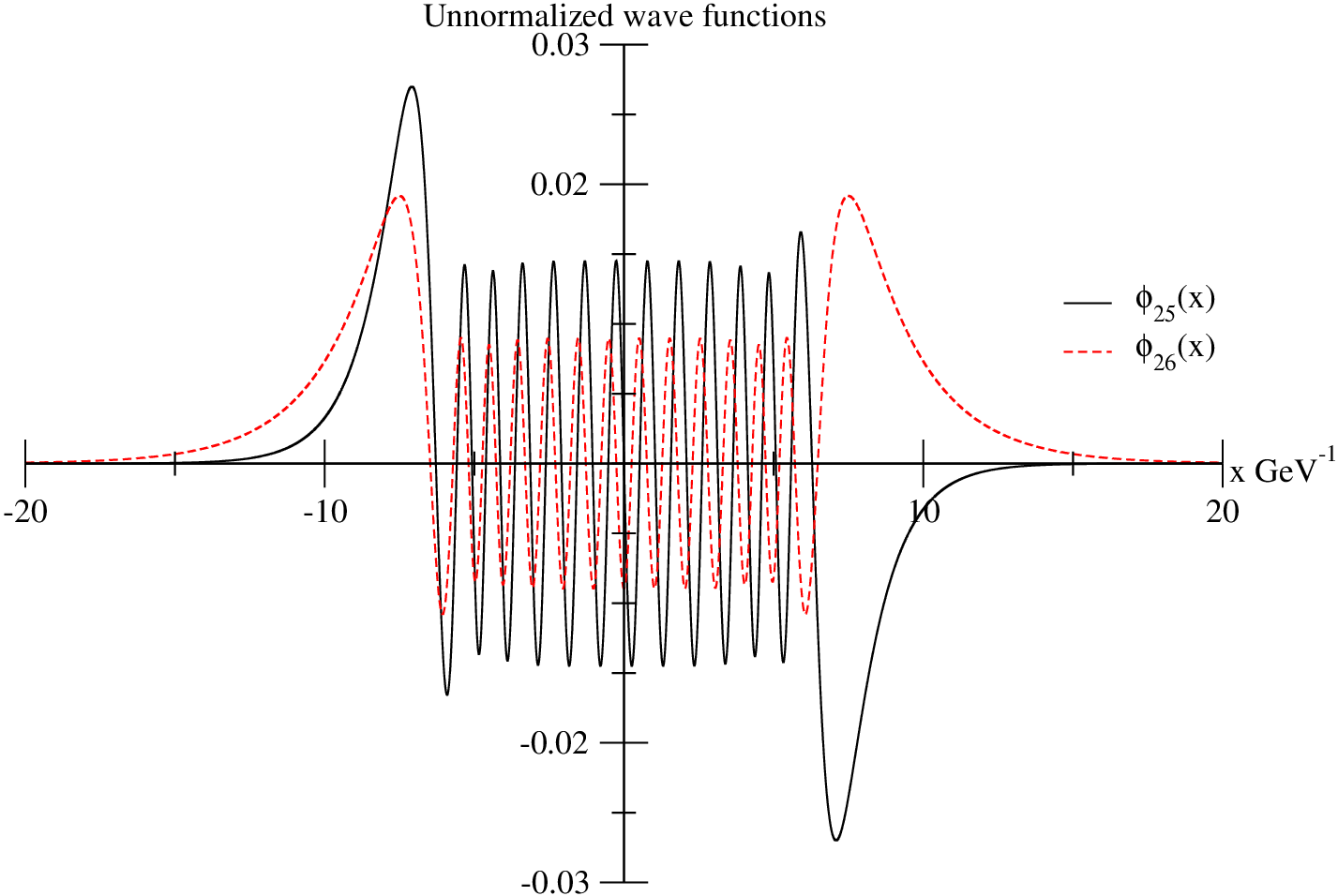}\\
  \caption{The last two unnormalized eigenfunctions.}\label{Boundstates-lasttwowave}
\end{figure}

We use the natural units system and take the mass of the confined particle as $2$ $GeV$ and calculate the energy spectrum. The obtained spectrum is given in  Table~\ref{boundeigenvalues} with the node numbers that are denoted with $n$. In Fig.~(\ref{Boundstates-threewave}) and Fig.~(\ref{Boundstates-lasttwowave}) We plot the corresponding unnormalized wave functions  to the eigenvalues $E_0$, $E_1$, $E_2$
and $E_{25}$, $E_{26}$, respectively.

\section{Conclusion \label{sec6:conclusion}}
We analytically investigate the scattering and bound state solution  of the KG equation with the multi-parameter q-deformed Woods-Saxon type potential energy under the spin symmetric limit in one spatial dimension. We prove the conservation of the total probability in the scattering case, after deriving the reflection and transmission probabilities. Then, we examine the correlation between the potential parameters with the reflection and transmission probabilities with assigning numerical values to the parameters of the potential barrier randomly. Moreover, we use the continuity conditions in the bound state case, and point out  to a quantization scheme to obtain an energy spectrum. Finally, we use the Newton-Raphson method to obtain an energy spectrum numerically. Therefore,  we report that  multi-parameter q-deformed Woods-Saxon type potential energy is an appropriate  candidate to explore the diatomic and polyatomic molecules structures in molecular physics.

\section*{Acknowledgments}
One of the authors, B.C. L\"utf\"uo\u{g}lu (BCL), was partially supported by the Turkish Science and Research Council (T\"{U}B\.{I}TAK) and Akdeniz University. BCL thanks for the support given by the Internal Project of Excellent Research of the Faculty of Science of University Hradec Kr\'{a}lov\'{e}, "Studying of properties of confined quantum particle using Woods-Saxon potential". We thank Prof. M. Horta\c{c}su for the proof reading. Finally, the authors thank the kind referees for their positive suggestions and comments to improve the quality of this manuscript.

\newpage

\begin{table}
 \begin{center}
\begin{tabular}{|c|c|c|c|c|c|c|c|c|c|c|c|}
  \hline
  $\tilde{\alpha} =\alpha $ &$\tilde{L}=L$ & $\tilde{V}_1=V_1 $ &$\tilde{V}_2 =V_2 $&    $\tilde{A}=A$ & $\tilde{B}=B$ &
    $\tilde{C}=C$  & $\tilde{D}=D$ & $\tilde{q} =q $ & $\tilde{p} = p$ &  $\tilde{\xi} =\xi $&  $\tilde{\eta}=\eta$\\
   \hline
    $2.0$  &$4.0$ &$1.0$&$0.2$ &$ 0.1$ & $ 1.0$& $ 0.1$ & $10.0$ &$0.8$ &$8.0 $ & $5.0$&$10.0$  \\   \hline
    $GeV$ &$\frac{1}{GeV}$&$GeV$& $GeV$                 &$GeV$&$GeV$&$\sqrt{GeV}$&$\sqrt{GeV}$ &$none$&$none$&$none$&$none$  \\   \hline
\end{tabular}
\caption {Arbitrarily assigned positive parameters in Natural units to build a symmetric potential energy function.}\label{parameters}
\end{center}
\end{table}

\begin{table}
 \begin{center}
\begin{tabular}{|c|c|c|c|c|c|c|c|c|c|c|c|c|c|c|c|c|c|}
  \hline
  $n$ & $E_n$  & $n$ & $E_n$  & $n$ & $E_n$ & $n$ & $E_n$  & $n$ & $E_n$  & $n$ & $E_n$ & $n$ & $E_n$  & $n$ & $E_n$  & $n$ & $E_n$ \\
   \hline
  $0$ & $-1.998$    & $3$ & $-1.874$ & $6$ & $-1.561$ &  $9$ & $-1.126$   & $12$ & $-0.611$ & $15$ & $-0.051$ &$18$ & $0.528$    & $21$ & $1.102$ & $24$ & $1.643$ \\
  \hline
   $1$ & $-1.979$    & $4$ & $-1.786$ & $7$ & $-1.428$ &  $10$ & $-0.961$ & $13$ & $-0.428$ & $16$ & $0.141$ &$19$ & $0.721$    & $22$ & $1.288$ & $25$ & $1.805$ \\
  \hline
  $2$ & $-1.939$    & $5$ & $-1.681$ & $8$ & $-1.282$ &  $11$ & $-0.789$  & $14$ & $-0.241$ & $17$ & $0.334$ &$20$ & $0.913$    & $23$ & $1.469$ & $26$ & $1.943$ \\
  \hline
\end{tabular}
\caption {Bound state energy spectrum. $n$ denotes the node number. Note that eigenvalues have units in GeV. } \label{boundeigenvalues}
\end{center}
\end{table}

\newpage

\section{References}

\newpage

\newpage

\newpage
\newpage

\newpage

\newpage

\newpage

\newpage

\newpage

\newpage

\newpage

\newpage

\newpage

\newpage
\end{document}